\documentclass[12 pt]{article}
\usepackage{mathptmx}
\usepackage{amssymb}
\usepackage{graphicx}
\usepackage[dvips]{color}
\usepackage[hyphens]{url}
\usepackage{hyperref}
\usepackage[german,english]{babel}
\usepackage{natbib}
\setcitestyle{round,aysep={},yysep={;},notesep={;~},citesep={;}}

\long\def\symbolfootnote[#1]#2{\begingroup%
\def\thefootnote{\fnsymbol{footnote}}\footnote[#1]{#2}\endgroup} 

\def\citeapos#1{\citeauthor{#1}'s (\citeyear{#1})}

\begin{document}
\title{Einstein's cosmological considerations}
\author{Daryl Janzen\thanks{email: daryl.janzen@usask.ca}} 
\date{University of Saskatchewan, Saskatoon, Canada}
%
%
%
\maketitle
\abstract{The objective of this paper is not simply to present an historical overview of Einstein's cosmological considerations, but to discuss the central role they played in shaping the paradigm of relativistic cosmology. This, we'll show, was a result of both his actions and, perhaps more importantly, his inactions. Accordingly, discussion won't simply be restricted to Einstein's considerations, as we'll analyse relevant contributions to the relativistic expansion paradigm during the approximately twenty years following Slipher's first redshift measurements in 1912. Our aim is to shed some light on why we think some of the things we do, with the idea that a better understanding of the reasoning that fundamentally influenced the common idea of our expanding universe might help to resolve some of the significant problems that modern cosmology now faces; and we eventually use this knowledge to probe the foundations of the standard model. Much of the information we present, including many of the historical details, we expect will be news to modern practitioners.} 

\section{Introduction}
\label{intro}

Cosmology is in a crisis state unprecedented in the history of science. For while we now possess a model describing the large-scale evolution of our Universe whose parameters have been constrained with significant precision, many aspects of the universe that this model seems to describe simply don't meet our theoretical expectations---and a number of `big questions' have therefore arisen along with the great scientific progress that's taken place this past century. In fact, the model that we have is a remarkably simple one, in which a well-behaved trigonometric function describing cosmic expansion is parametrised effectively with just a single non-trivial value. Particularly in the last fifteen years, this one parameter-model has been tested against the data and various physically realistic complications that could be made to it, passing all tests as the one that statistically provides the best fit to our observations. Our Universe, it would seem, is an extraordinarily simple one. 

The standard model accurately \textit{describes} the cosmological data, yet the nature of what we're measuring remains \textit{unexplained}: we lack good reasons why observations should be as they are; why we should \textit{expect} to see what we do. In contrast to previous instances such as the Scientific Revolution, when the Ptolemaic model was replaced by Kepler's more accurate description, we now lack the guidance of discrepancy between model and observation that science is fit to resolve, and have only discontent with an impossibly fine-tuned description of the facts, to lead us in the next step. 

The difference between description and explanation is mostly a difference between science and philosophy; and unfortunately this time science seems to have taken us as far as it's able, reaching a dead end in constraining an unexpected description. Indeed, what we now lack is a good theoretical argument for \textit{why} the Universe should be as it is so well described; for while the description that's been constrained is a particular case that's \textit{allowed} by our general model structure, the reason why this specific model \textit{should} be favoured in Reality has eluded us. 

In fact, not only have many of our empirical results been unexpected, but often our expectations were that they \textit{shouldn't} have been as they are---that they really \textit{should} have been otherwise. The most telling example of this discrepancy between expectation and observation is found in a paper by \citet{Einstein1932}. There, the cosmological-term in the field equations is tossed out right away, as ill-motivated and useless, while the curvature of space is set to zero as a simplifying assumption, with a note that its actual sign and value would eventually be fixed through more precise data. 

Later, \citet{Einstein1945} noted that once we take the idea of an expanding universe seriously, ``the main question becomes whether space has positive or negative spatial curvature.'' Precise flatness, in an Einstein-de~Sitter universe, is totally unexpected: curvature is supposed to be due to the density of world-matter, so precise flatness would be an infinitely unlikely, fine-tuned possibility. In fact, the smallest deviation from flatness after the big bang should theoretically have blown up to an enormous one by now, so at the very least the sign of the curvature should be obvious, if not its value. 

In contrast to the impossibility (or: infinitely unlikely possibility) that zero spatial curvature would actually be observed, the cosmological-term was to be rejected as superfluous and unjustified---or equivalently, the value of the ``cosmological constant'' $\lambda$\footnote{The cosmological constant has for decades been commonly written with an upper-case $\Lambda$ rather than the lower-case $\lambda$ that was used initially. However, because our discussion mostly concentrates on early work, the majority of our quotations predate the switch, so we use the lower-case in the main text as well, for general consistency.}  was supposed to be precisely zero.

But eighty years of experimental progress has shown exactly the opposite: increasingly refined measurements constrain our model parameters ever closer to a value of zero curvature \citep{Hinshaw2013,Ade2013}---and, since the turn of the century \citep{Riess1998,Perlmutter1999}, we've come to realise that, in all likelihood, $\lambda$ is positive \citep{Riess2004,Riess2007,Davis2007,Hicken2009,Riess2011,Suzuki2012,Hinshaw2013,Ade2013}.

Through the course of scientific progress and discovery, additional features were put into the picture in an attempt to make the unexpected observations expected---to dress up our generic model and make it look like the vision of perfection that we observe. The inflation scenario was proposed in part to account for the so-called ``flatness problem'' just mentioned \citep{Guth1981}, and since the discovery of accelerated cosmic expansion, it's been hoped that the driving force behind that would come from a \textit{quintessence} field that also drove inflation early on; see, e.g., \citet{Weinberg2008} for a review. 

However, inflation may not be up to the task \citep{Steinhardt2011}. In a recent critical assessment of the inflation scenario's effectiveness, \citet{Ijjas2013} considered that ``In testing the validity of any scientific paradigm, the key criterion is whether measurements agree with what is expected given the paradigm''; and indeed, their results showed that the inflation scenario, which was originally invented as a way \textit{out} of needing to fine-tune the Universe, has to be fine-tuned \textit{itself} in order to agree with the most recent data. Whereas many physicists remain undeterred, Steinhardt, who was one of the theory's originators, thinks ``the field is over-enamored with inflation---unable to see its serious foundational flaws and now its troubles with the new data'' (private correspondence). 

Perhaps an alternative approach to dealing with the fact that little of what we observe is really expected---i.e. rather than beginning with a universe that ideally shouldn't be at all like ours \textit{actually} is, and trying to `dress it up' so that it looks the way ours does---is to re-examine why we think it should fundamentally be something so different from the way it looks. If our basic problem is that we don't expect anything we see, then it may well be that our expectations are wrong---and a solution to our current problems could then be found by taking a closer look at the foundations of our theory and looking to revise them. 

In the hundred years since relativistic cosmology emerged as a discipline, a lot of new, and often unexpected, facts have been uncovered. It seems that by re-examining the course of reasoning that produced the theory upon which our current expectations are based, and keeping the presently known facts of observation in mind, we might find that our expectations really should be different. By understanding the intricacies of the reasoning that produced the standard model---by really knowing why it's thought to represent what it is, \textit{and why we're supposed to want it to do that}---we might see why we should want cosmology to represent something else---something through which the empirical evidence basically \textit{should}, without needing to be `dressed up', be what's \textit{expected}. 

Indeed, there is an interesting and very revealing story to be told about the origins of the current cosmological paradigm, which explains the basic reasoning that stands behind our current expectations. In particular, as we examine the history we'll see that there were key debates that should have taken place in regard to the nature of cosmic expansion, which were never definitively resolved, either from a theoretical standpoint or as a result of improved empirical knowledge. And indeed, it now appears that theoretical perspectives which were marginalised or simply got neglected in the course of scientific progress may be the best suited to explain the number of unexpected facts that have since been observed---whereas the choices that traditionally came to be accepted without question have stood directly opposed to any possible explanation of the phenomena.

Our investigation of modern cosmology's foundations therefore begins with a detailed examination of the history of thought in relativistic cosmology's developmental stage, concentrating on the first twenty years or so, following the beginning of Slipher's campaign to catalogue the Doppler shifts of spiral nebulae \citep{Slipher1913}. Einstein, as the founder of the theory of gravitation on which modern theoretical cosmology was based, and the crafter of the first relativistic cosmological model, had a particularly prominent and highly influential role in the early development of relativistic cosmology, and therefore takes centre stage in our discussion. However, because he hardly wrote anything on the cosmological problem after 1917, we must necessarily supplement the brief but influential statements he did make afterwards, with a broader discussion of what was going on at the time. For while Einstein's opinions in regard to the cosmological problem have always been given due consideration, the basic reasoning behind them, and even his main aims, have often remained somewhat nebulous, and relating his opinions to the contemporary considerations of others will sometimes give the clearest indication of what his later reasoning was.

In a way, much of the thrust of our historical account aims to understand the details behind his rejection of the cosmological-term, and why he considered its introduction ``the biggest blunder he ever made in his life'' \citep{Gamow1970}. This expression will be clarified through an historical reconstruction of his evolving preference of particular cosmological models in light of (experimental and theoretical) scientific progress. As we'll see, Einstein's rejection of the term, which had previously been thought to describe a repulsive force that would naturally drive cosmic expansion, was a significant turning point in theoretical cosmology. Indeed, \citeapos{Hubble1929} `discovery' that the Universe is expanding was very much expected from a theoretical standpoint---as the linear redshift-distance relation, which was qualitatively supported by redshift measurements, had been prominently considered already since the early 1920s. 

There's good evidence that, already in 1923, Einstein had favoured \citeapos{Friedmann1922} original closed universe solution with $\lambda=0$, but awaited Hubble's \textit{empirical confirmation} of expansion before formally announcing his preference. On the other hand, the evidence shows that the papers by both \citet{Friedmann1922,Friedmann1924} and \citet{Lemaitre1927} remained unknown to other theorists who were more willing to speculate about the expansion problem, and had taken an interest in the relativistic description of potentially systematic redshifts in particular, in the early 1920s. Much in our account of this early work will likely be news to those who have thought that ``cosmologists (theoreticians and observers alike) were convinced that the universe must be static, and so\ldots they resisted the idea of an expanding universe for about a decade'' until the spring of 1930 \citep{Ellis1999}.

Indeed, there has been a great deal of confusion surrounding the manner in which the idea of an expanding universe historically came to be accepted. \citeapos{Hubble1929} discovery is often presented as a turning point in relativistic cosmology, which ``forced'' the concept of expansion on people who would rather have thought the Universe should be static, and therefore weren't open to the few more adventurous proposals of expanding cosmological models by \citet{Friedmann1922,Friedmann1924} and \citet{Lemaitre1927}---and this simply isn't true. But what's even more troubling to find in the historical accounts that are lately being written by physicists, philosophers, and popular science writers alike, is that people have started to believe the reason why Einstein regretted the introduction of the $\lambda$-term is that he could otherwise have been the one to predict the expansion of the Universe, which isn't at all what was on his mind. The irony of this growing misconception is that, to \citet{Eddington1933}, Einstein \textit{did} get credit for ``almost inadvertently'' predicting the expansion of the universe \textit{because} he added the term.

But in the years since Hubble's discovery, the concept of how expansion occurs has changed a great deal from what Eddington and others had in mind at the time. Decades ago already, \citet{Gribbin1986} painted a very different picture of the discovery of the expanding universe than what actually took place. There, after misattributing Hubble's role as the discoverer of the very large velocities of distant galaxies,\footnote{Credit for this discovery is owed to Slipher, who was the first to measure the ``velocities'' of the spiral nebulae (now recognised as galaxies), and the first to sufficiently demonstrate that they are predominantly (though not, as Gribbin tells us, ``all'') receding. Hubble, on the other hand, discovered a means of determining their distances, and subsequently demonstrated a \textit{systematic redshift-distance relation}, which was the first concrete piece of evidence that our Universe must be expanding.}  he added, ``The reason for the existence of the lambda term had gone, and the existence of the expanding Universe remains the greatest prediction Einstein never made'', before going on to quote Einstein as describing ``the introduction of the lambda term as `the biggest blunder of my life'''.

More recently, \citet{Dainton2001} wrote that, as a result of Hubble's discovery, ``Einstein embraced the expanding model of universe [sic], and repudiated the cosmological constant -- calling it his `greatest mistake' -- not surprisingly regretting having missed the opportunity of going down as discoverer of the dynamic cosmos.'' And \citet{Penrose2005} has also stated that ``when it became clear, from Edwin Hubble's observations in 1929, that the universe is expanding, and therefore \textit{not} static, Einstein withdrew his support for the cosmological constant, asserting that it had been `his greatest mistake' (perhaps because he might otherwise have predicted the expansion of the universe!)''. 

A number of facts are misconstrued in these last two statements. First of all, there's no record that Einstein ever referred to the cosmological constant as his ``greatest mistake'': the statement likely originates from the same one loosely quoted by Gribbin, from \citeapos{Gamow1970} posthumously published autobiography, where he alleged that Einstein told him in conversation much later that the introduction of the $\lambda$-term was the ``biggest blunder'' he ever made in his life. In fact, it's reasonable to guess that the ``greatest mistake'' mis-quote by both of these authors originates from Gribbin's remark that the expansion of the Universe is ``the greatest prediction Einstein never made''---i.e. effectively that it was Einstein's ``greatest mistake'' to miss out on predicting expansion because he introduced the $\lambda$-term instead; see also \citet{Kolb2007}. 

Now,---more to the point that the statements by Dainton and Penrose misconstrue the facts---the reason why, as \citet{Gamow1970} tells us, Einstein eventually  ``remarked that the introduction of the cosmological term was the biggest blunder he ever made in his life'', was really already suggested by Gamow, who indicated that Einstein had really wanted the term stricken from general relativity, but once introduced it just wouldn't go away. Furthermore, while this comment certainly fits with other statements made by Einstein, it was made privately \textit{and} much later than the actual discovery of the expanding universe; he didn't immediately withdraw $\lambda$ with a proclamation that his `greatest mistake was to introduce the term, and thereby miss out on predicting cosmic expansion,' as all three authors suggest. In fact, if Einstein had stated his actual opinion so overtly earlier on when he rejected the term, people would have known better how to take him. But his rejection of the cosmological constant was always either too cryptic\footnote{For example, \citeapos{Einstein1931a} famous rejection of the term asked whether the facts can be accounted for without the ``anyhow theoretically unsatisfactory $\lambda$-term'', then showed that they can, and finally concluded by stating that ``Above all, it is remarkable that Hubble's new facts allow general relativity theory to seem less contrived (namely, without the $\lambda$-term)\ldots''---without ever offering an actual reason why the term was theoretically unsatisfactory or why it made the theory seem contrived. 

\indent{In this, his second paper on the cosmological problem of relativity theory, written fourteen years after he had carefully justified the introduction of the $\lambda$-term, the fact that it was at least theoretically unsatisfactory and made general relativity theory seem contrived, was something he just took for granted. His famous rejection of the term he had previously taken great care to introduce, was simply \textit{adjectival}!}} or his arguments were just so weak\footnote{For example, a repeated argument he made was that the $\lambda$-term was introduced for the purpose of constructing a static model, and wasn't necessary in particular non-static cases, so it lost its justification \citep{Einstein1932,Einstein1945}. \citet{Lemaitre1949a} actually openly attacked this argument in an essay written for Einstein, noting that ``Even if the introduction of the cosmological constant `has lost its sole original justification, that of leading to a natural solution of the cosmological problem,' it remains true that Einstein has shown that the structure of his equations quite naturally allows for the presence of a second constant besides the gravitational one. This raises a problem and opens possibilities which deserve careful discussion. The history of science provides many instances of discoveries which have been made for reasons which are no longer considered satisfactory.'' 

\indent{Einstein's response to this criticism was simply to throw his old argument right back at Lema\^{i}tre \citep[p.~685]{Einstein1949}: ``After Hubble's discovery of the `expansion' of the stellar system, and since Friedmann's discovery that the unsupplemented equations involve the possibility of the existence of an average (positive) density of matter in an expanding universe, the introduction of such a constant appears to me, from the theoretical standpoint, at present unjustified.''}}  that more than anything his stance confused and frustrated people like Weyl, Tolman, Eddington, and Lema\^{i}tre.

This brings us to the particular Friedmann model that \citet{Einstein1931a} eventually showed a preference for: a 3-sphere that expands from a singularity out to a maximum radius, before collapsing back in symmetrical fashion. An air of uncertainty has always surrounded his choice, the reason for which is particularly confounding in light of the fact that he cast it off the very next year, in favour of an eternally expanding model \citep{Einstein1932}. \citet[p.~177]{Steinhardt2007} suggest some reasons why Einstein would have favoured the model of a periodic oscillatory universe over \citeapos{Friedmann1924} open universe solution, and indicate that he gave up on his original hopes only when he found that the age of the closed universe would have to be less than the age of the Earth. 

However, we'll see that Einstein \textit{didn't} actually favour the one Friedmann model over the other, and his choice in 1931 likely had nothing to do with the possible periodicity of the ``oscillatory'' universe. And, since the so-called ``time-scale difficulty'' \citep{Bondi1960} was actually a major problem for the $\lambda=0$ Friedmann models with \textit{general} curvature before \citet{Sandage1958} finally revised the measured value of Hubble's constant, this can't have been the reason why Einstein changed his tune the next year. 

In fact, Einstein only favoured the particular Friedmann model he did in 1931 because, at the time, it was the one model he knew of that allowed him to reject the cosmological constant---which was his sole purpose in writing the paper \citep{Einstein1931b}. \textit{He didn't know about \citeapos{Friedmann1924} second solution, and his support for the ``oscillatory'' cosmology was purely coincidental}. And the reason he so quickly changed his mind when he published the article with de~Sitter in 1932, opening his support to the models with general spatial curvature---but still with $\lambda=0$!---was that \textit{\citet{Heckmann1931}} had independently rediscovered the general solutions with arbitrary curvature and $\lambda$ in the meantime---so Einstein hastened to show his preference once more to abandon the $\lambda$-term, allowing for \textit{all} solutions where that could be. 

While things are certainly clarified from a modern perspective by observing that Einstein hadn't known about \citeapos{Friedmann1924} second paper in the wake of Hubble's discovery, it's interesting to note that even his peers---who were just as unaware of \citeapos{Friedmann1924} second paper as he was---seem not to have fully appreciated that the \textit{only point} of his 1931 paper was to dump the $\lambda$-term in light of Hubble's empirical result, which provided the empirical justification to do so via the special case in \citeapos{Friedmann1922} solution. 

Indeed, soon after \citeapos{Einstein1931a} paper was published, \citet{deSitter1931} quickly wrote a note on the ``oscillating solution'' Einstein had shown his preference for, which he had neglected to pay special attention to in previous work; however, de~Sitter paid no special attention to Einstein's rejection of the $\lambda$-term, and simply compared the oscillating model's behaviour to that of the eternally expanding solutions (with positive curvature) and ``Einstein's [static] world'', with $\lambda>0$.

Tolman's response was equally off the mark. The two had met earlier that year during Einstein's first visit to Pasadena, and Einstein sent Tolman a letter that July, along with a copy of the new paper. In his response, Tolman took the ``quasi-periodicity'' of the model as his main point of interest, on which he had in fact followed up with a paper of his own \citep{Tolman1931b}---while he actually argued that setting $\lambda=0$ seemed arbitrary and not necessarily correct, even though \citet{Einstein1931b} opened his letter by saying that was the only point of the paper. 

Even much later, \citet{Lemaitre1949a} seems to have thought that the point was up for debate with Einstein, and attempted to invoke the $\lambda$-term for reasons quite similar to the ones Einstein had originally given (and likely not realising just how much he regretted that). As mentioned above, Einstein's response was simply as it had always been---\textit{for his sole purpose in regarding the cosmological problem of relativity from 1931 onwards} (and arguably from 1923 onwards) \textit{was to reject the cosmological constant}. 

This point, which was often mistaken even by friends and colleagues, will be brought out here by piecing together the evidence from Einstein's published work and private correspondence, and noting other influential factors such as his ignorance of \citeapos{Friedmann1924} second cosmology paper in the wake of \citeapos{Hubble1929} discovery. Thus, the blinds will be drawn back so we can have a good look at the original reasons that led to the current paradigm in cosmology, in which nothing but a faint 13.8 billion year old relic radiation field has really turned out to be as it was expected.  

We'll find that perhaps Einstein's most obvious impact on cosmology, due to his unwavering rejection of the cosmological constant, was to completely reverse the manner in which cosmic expansion was once conceived as a theoretical prediction---an expectation due to the way gravity was described to work at great distances, that differentiated general relativity no less from Newtonian mechanics than its description of Mercury's perihelion advance had \citep[pp.~24-25]{Eddington1933}. But every bit as significant as this, was his influence on the way cosmologists would always conceive of time, describing it---completely against the spirit of relativity!---as \textit{absolute}.

The latter point will not take the forefront in our discussion, just as it didn't take the forefront in Einstein's own considerations; but it will be addressed:---beginning with his initial assumption of a cosmic time that, while described as empirically Lorentz invariant, was no less absolute than the time of Newtonian mechanics; then in the reactions from his critics; and later, as it became a foundational assumption of cosmology, made again by him and everyone else who worked to develop the expanding universe theory; and finally, as we discuss the many degrees of its empirical confirmation, and possible ways forward from the significant deadlock in which we currently find ourselves. 
\section{Einstein and de~Sitter}
\label{sec:2}
Einstein clearly laid out the reasoning behind his cosmological model of 1917, presenting it in a paper entitled ``Cosmological Considerations on the General Theory of Relativity'' \citep{Einstein1917}. There is no mystery about why he chose to describe the Universe, at that time, as a static 3-sphere, assuming, against the spirit of relativity, that ``There is a system of reference relatively to which matter may be looked upon as being permanently at rest''. He considered both this ``approximative assumption'' and the assumption that the Universe is statical, to be motivated by empirical data: ``The most important fact that we draw from experience as to the distribution of matter is that the relative velocities of the stars are very small as compared with the velocity of light.''

Indeed, in a special relativistic world, one should expect to observe neighbouring bodies with velocities drawn uniformly from the interval (0, $c$), so it was reasonable for Einstein to infer from the evidence, that the world lines of all the stars in the known universe are parallel to each other, and perpendicular to the three-dimensional world in which they remain approximately at rest. And while it would turn out that better empirical evidence actually stands against the statical assumption, when Hubble showed twelve years later that there is a redshift-distance relation that holds for distant spiral nebulae, this later result was interpreted as resulting from the expansion of three-dimensional space, with the distant nebulae similarly all remaining roughly at rest as the Universe expands. Actually, this particular assumption of Einstein's has stood the test of time, as the redshift-distance relation has now been confirmed out to redshifts greater than 2, and we've observed thousands of objects with redshifts up to more than 8; for if these redshifts were due to relative velocities through space, rather than the transmission of light through an expanding three-dimensional universe, the relativistic speed limit tells us they couldn't exceed 1.

It's therefore now clearer than ever, both from these coarse-grained measurements and from detailed cosmic microwave background observations (as we'll discuss further in \S\,\ref{sec:6}), that we live in a three-dimensional Universe that expands in the course of cosmic time---and that the proper motions of galaxies really are ``very small as compared with the velocity of light'', much as Einstein recognised nearly a century ago.

Another assumption that Einstein made in constructing his model was that the Universe should be spherical---i.e. unbounded, but finite in extent. This assumption was made as the result of a realisation that sufficient boundary conditions for the gravitational potential at infinity could not be formulated. In fact, this is the first problem Einstein considered in the paper---and he considered it in great detail, `conducting the reader over the rather rough and winding road that he himself travelled, in hopes that a greater interest would be taken in the result at the end of the journey' (to paraphrase his remark). The result of all of this, was that he would slightly modify his gravitational field equations, adding the (in)famous ``cosmological-term''.

\citet{Eddington1933} eventually commented: 
\begin{quotation}
Einstein made a slight amendment to his law to meet certain difficulties that he encountered in his theory. There was just one place where the theory did not seem to work properly, and that was---infinity. I think Einstein showed his greatness in the simple and drastic way in which he disposed of difficulties at infinity. He abolished infinity. He slightly altered his equations so as to make space at great distances bend round until it closed up. So that, if in Einstein's space you keep going right on in one direction, you do not get to infinity; you find yourself back at your starting-point again. Since there was no longer any infinity, there could be no difficulties at infinity. \textsc{q.e.d.}
\end{quotation}
Eddington's opinion was not, however, universally held; e.g., when he published this statement, Einstein had already rejected the cosmological-term and the finiteness of space, along with the static assumption. In fact, years before Hubble found his famous linear redshift-distance relation, Einstein was influenced by the work of de~Sitter, with whom he had been in contact at the time he published his ``Cosmological Considerations''. \citeapos{Einstein1917} paper was published in February, and just a month later, \citet{deSitter1917a} published a response that basically countered every one of his cosmological considerations.

In a few short sentences, de~Sitter had cast off Einstein's concerns regarding the boundary conditions, as invalid: first of all, from an experimental point of view, as something impossible to ever observe anyway; and then from a theoretical (mathematical) point of view, as he showed with ease, given the new $\lambda$-term, that the ``most desirable and the simplest'' possible boundary constraint---viz. all $g_{\mu\nu}$ at infinity equal to zero---was indeed achievable. But what made matters worse was that de~Sitter's world contained none of the ``hypothetical world-matter'' that Einstein thought should be essential to any general relativistic model---an assumption which, de~Sitter noted, ``serves no other purpose than to enable us to suppose it not to exist.'' 

Further on in the paper, de~Sitter attacked the boundary values again, asking which of the three systems---Einstein's, labelled \textit{A}; his new solution, labelled \textit{B}; or the Minkowski metric of special relativity, labelled \textit{C}---should be preferred. His answer: ``From the purely physical point of view, for the description of phenomena in our neighbourhood, this question has no importance'': space-time is Minkowskian and the amended field-equations reduce to the old ones ``in all cases within the limits of accuracy of our observations''. Therefore, he pointed out, the choice of how we would extrapolate outside our neighbourhood cannot be decided by physical arguments, ``but must depend on metaphysical or philosophical considerations, in which of course also personal judgment or predilections will have the same influence.''

He repeatedly criticised Einstein's assumption concerning the division between time and space, e.g. stating most significantly that in Einstein's solution ``time has a separate position'' from space, which
\begin{quote}
is evident a priori. For speaking of \textit{the} three-dimensional world, if not equivalent to introducing an absolute time, at least implies the hypothesis that at each point of the four-dimensional space there is one definite coordinate $x_4$ which is preferable to others to be used as ``time'', and that at all points and always this one coordinate is actually chosen as time. Such a fundamental difference between the time and the space-coordinates seems to be somewhat contradictory to the complete symmetry of the field-equations and the equations of motion (equations of the geodetic line) with respect to the four variables.
\end{quote}

Then, in a postscript, de~Sitter addressed Einstein's concern that ``the field should be due to the matter, without which it cannot exist'', which Einstein had communicated to him privately. He wrote that if Einstein would choose this ``postulate of the logical impossibility of supposing matter not to exist'', that would indeed be satisfied only by choosing Einstein's solution; but, he noted, this comes with the price of ``introducing the constant $\lambda$, and assigning to the time a separate position amongst the four coordinates.'' On the other hand, he wrote, his own solution satisfies the postulate that the $g_{\mu\nu}$ should be invariant at infinity---noting once more, however, that this guiding principle of Einstein's cosmological considerations ``has no real physical meaning''. He then added that there is complete relativity of time in his solution, but that it also needs to assume the $\lambda$-term. 

The paper concludes by noting that both systems \textit{A} and \textit{B} can be rejected, and that we can stick with the original field equations and the Minkowski metric, which is not invariant at infinity. But then, he notes, inertia is not explained: ``we must then prefer to leave it unexplained rather than explain it by the undetermined and undeterminable constant $\lambda$. It cannot be denied that the introduction of this constant detracts from the symmetry and elegance of Einstein's original theory, one of whose chief attractions was that it explained so much without introducing any new hypothesis or empirical constant.''

de~Sitter tidied up the main parts of this paper and included them, along with some additional calculations, in his seminal ``Third Paper'' in the Monthly Notices of the Royal Astronomical Society the following November \citep{deSitter1917b}. His argument against Einstein's introduction of the $\lambda$-term fell through the cracks, and for years Einstein's original model was contrasted with de~Sitter's, as the two main competing cosmological solutions, both with $\lambda>0$.

However, it seems that this criticism of his deep considerations, and particularly of the new $\lambda$-term, from a friend he greatly respected, had an effect on Einstein. For instance, it's notable that the 1917 paper was the last one where he published the details of his cosmological considerations. In fact, even following Hubble's discovery, which whipped the entire community into a frenzy, Einstein was nearly mute, publishing a total of five pages on the problem, in two separate articles that promoted two different models---the second one even being co-authored by de~Sitter. In contrast, de~Sitter published volumes on the matter in the few short years he had left. 

It may be that Einstein had simply lost his taste for the cosmological problem, with so much else going on during that era---especially given de~Sitter's prompt negative response to his previous considerations. Whatever the reason for Einstein's lack of voice later on, the two papers he did write carried enormous weight---so much so that the two-page paper written with de~Sitter was still looking to be confirmed at the turn of the century, when it was discovered instead that the Universe is presently accelerating in its expansion \citep{Riess1998,Perlmutter1999}. In many ways, Einstein's \textit{later} considerations, and particularly his rejection of the cosmological constant, helped shape the paradigm for cosmic expansion in the latter half of the twentieth century. But in order to appreciate that, it's important to begin by forming a better idea of the climate in which Einstein's later opinions were disseminated, by examining the work that further influenced his considerations.

\section{The evidence for expansion}
\label{sec:3}
Einstein's 1917 paper on cosmology was his last primary contribution to the field. All subsequent theoretical advances on the cosmological problem were made by others, and it was through their results that Einstein formed his later opinions. Following the episode with de~Sitter, Einstein's cosmological considerations were significantly influenced by the works of \citet{Eddington1923}, \citet{Weyl1923a,Weyl1923b}, and \citet{Friedmann1922}. 

In the early 1920s, Eddington and Weyl in particular were engaged developing de~Sitter's solution as the relativistic description of an expanding universe, and sought support from the redshift measurements from spiral nebulae (which were later understood to be galaxies like our Milky Way). Because it was the redshift observations that motivated their investigations, it's worth briefly recounting that history. In 1912, \citet{Slipher1913} began a campaign to measure the relative radial velocities of spiral nebulae from Doppler shifts in their spectra. His stated purpose was that, while previous work seemed to ``suggest that the spiral nebula is one of the important products of the forces of nature'', which had already led to extensive photographic study of their spectra, no attempt had yet been made to determine their radial velocities, ``although the value of such observations has doubtless occurred to many investigators.'' 

Thus, the first piece of evidence that contributed to the greatest change in our understanding of the cosmos since the Scientific Revolution---a discovery every bit as important to the advance of human knowledge as the discoveries of the moons of Jupiter and the Lunar landscape---wasn't based on much more than a simple shot in the dark. For what Slipher discovered was that ``the Andromeda Nebula is approaching the solar system with a velocity of about 300 kilometers per second.'' Thus, he concluded: ``That the velocity of the first spiral observed should be so high intimates that the spirals as a class have higher velocities than do the stars and that it might not be fruitless to observe some of the more promising spirals for proper motion.''

Now, de~Sitter was certainly aware of the redshift measurements by Slipher and others that had accumulated by 1917. And what's more: he had found, in his new solution (which he thought described a statical world), that the spectra of distant objects would be systematically redshifted. Therefore, when he wrote his third paper ``On Einstein's Theory of Gravitation, and its Astronomical Consequences'' \citep{deSitter1917b}, he noted the radial velocities of three spiral nebulae that had already been measured by more than one observer: Andromeda's, at -311~km/s, had been measured by three observers; NCG 1068, at +925~km/s, had also been measured by three observers; and NGC~4594, at +1185~km/s, had been measured by two observers. He interpreted the observations as follows \citep{deSitter1917b}: 
\begin{quote}
The velocities due to inertia\ldots have no preference of sign. Superposed on these are, however, the apparent radial velocities due to the diminution of $g_{44}$, which are positive. The mean of the three observed radial velocities stated above is +600~km/sec\ldots Of course this result, derived from only three nebulae, has practically no value. If, however, continued observation should confirm the fact that the spiral nebulae have systematically positive radial velocities, this would certainly be an indication to adopt the hypothesis B [de~Sitter's] in preference of A [Einstein's].''
\end{quote}

While de~Sitter therefore didn't suggest an expanding solution, his interpretation is particularly interesting in the way it developed on a remark by \citet{Slipher1913}, that ``The magnitude of this velocity, which is the greatest hitherto observed, raises the question whether the velocity-like displacement might not be due to some other cause, but I believe at present we have no other interpretation for it.'' de~Sitter had now found an answer to Slipher's speculation---which was, incidentally, the first attempt to explain the observed redshifts relativistically, rather than as Doppler shifts due to relative proper motion. And it may have been thus, through its new-found use in accounting for the empirical evidence, that de~Sitter's solution was promoted, after its original presentation as the crux of an argument against Einstein's introduction of the $\lambda$-term, to a bona fide cosmological model. 

It wasn't long, though, before people realised that the solution doesn't describe a static universe, but one in which geodesics tend to separate from each other; and so the solution came to serve as the model of an expanding universe, which the redshift data seemed to suggest. 

The first attempts to interpret de~Sitter's solution as the model of an expanding universe were made by Eddington and Weyl; e.g., \citet{Eddington1923} considered the statical spherically symmetric line-element for de~Sitter space,
\begin{equation}
ds^2=-\frac{dr^2}{1-{1\over 3}\lambda{r^2}}-r^2d\theta^2-r^2\sin^2\theta{d\theta}^2+\left(1-{1\over 3}\lambda{r^2}\right)dt^2,
\end{equation}
from which he could write the equations of motion and show that geodesics do not remain at rest at constant radial distances. He went on: 
\begin{quotation}
Thus a particle at rest will not remain at rest unless it is at the origin; but will be repelled from the origin with an acceleration increasing with the distance. A number of particles initially at rest will tend to scatter, unless their mutual gravitation is sufficient to overcome this tendency.

It can easily be verified that there is no such tendency in Einstein's world\ldots It is sometimes urged against de~Sitter's world that it becomes non-statical as soon as any matter is inserted in it. But this property is perhaps rather in favour of de~Sitter's theory than against it.

One of the most perplexing problems of cosmogony is the great speed of the spiral nebulae. Their radial velocities average about 600 km. per sec. and there is a great preponderance of velocities of recession from the solar system.
\end{quotation}
And here, he produced a table of 41 spiral nebulae with radial velocity measurements that was obtained from Slipher, which was the most complete list as of February 1922. Of these, only five were negative (approaching). Then, after moving on to discuss ``Einstein's cylindrical world'', he concluded: ``Einstein's world offers no explanation of the red shift of the spectra of distant objects; and to the astronomer this must appear a drawback.'' 

At the same time, Weyl also engaged in the application of de~Sitter's solution to the cosmological problem, both in the fifth (and final) edition of \textit{Raum, Zeit, Materie} \citep{Weyl1923a} and in a follow-up paper \citep{Weyl1923b}. Weyl's work represents a significant advance on de~Sitter's and Eddington's, for a couple of reasons. First and foremost, Weyl recognised that in order for de~Sitter's solution to constitute a proper cosmological model, an assumption needs to be made about the state of rest of stars, much as \citet{Einstein1917} had realised when he constructed his model. 
	
As we've discussed above, \citet{deSitter1917a} had chastised this assumption of Einstein's, noting that, ``if not equivalent to an absolute time\ldots Such a fundamental difference between the time and space-coordinates seems to be somewhat contradictory to the complete symmetry of the field-equations and the equations of motion\ldots with respect to the four variables.'' In fact, the assumption is equivalent to an absolute time, as it defines an absolute (cosmic) state of rest and an absolute (cosmic) simultaneity-relation. The reason for de~Sitter's non-committal ``if not equivalent\ldots'' may be attributed to the fact that, despite this definition, the physics remains Lorentz covariant. Even so, in order to describe ``space'' that expands through the course of ``time'', the assumption that there is a true cosmological time has remained a basic axiom of cosmology. 

When \citet{Robertson1933} made this assumption in deriving the general line-element for cosmology, he attributed it to Weyl:
\begin{quote}
That we require the spatial distribution of matter in our highly idealized universe to be uniform implies the existence of a significant simultaneity, and would at first seem contrary to the postulates of relativity, according to which each observer refers the world to his own proper space and time\ldots That some assumption concerning the natural state of motion of the matter in the universe is required in order to account for the facts has been emphasized above all by Weyl.
\end{quote}
Further along in this paper, he referred to the assumption as ``Weyl's postulate'', a term which was prominently adopted by \citet{Bondi1960} in his popular \textit{Cosmology} book. Indeed, Weyl did much to formalise the principle relativistically; and, as Bondi noted, ``However much this may contradict orthodox relativity, the concept of a preferred direction is common to all cosmological theories.'' More formally, Bondi summarised Weyl's postulate as a statement that ``The particles of the substratum (representing the nebulae) lie in space-time on a bundle of geodesics diverging from a point in the (finite or infinite) past'', before going on to discuss its significance.

On the other hand, it should be noted that ``Weyl's postulate'' \textit{originated with \citet{Einstein1917}}, who was indeed conscious of the assumption he made, and even bore some flak for doing it. In fact, despite \citeapos{deSitter1917a} criticism \textit{and the fact that the assumption does stand opposed to the spirit of relativity}, describing the entire Universe as ultimately only \textit{empirically} Lorentz invariant, Einstein seems never to have wavered in his choice, as he'd eventually favour the descriptions in which isotropic and homogeneous space expands in cosmic time from an initial singularity, and without a cosmological constant \citep{Einstein1931a,Einstein1932,Einstein1945}.
	
The significance of this assumption for Weyl was that it allowed him to calculate the redshift of light from distant objects in de~Sitter space-time. He made the assumption as follows \citep{Weyl1923b}: ``That\ldots two stars\ldots belong to the same system and are causally connected through a common origin, implies that their world lines have the same action domain $\Sigma$'', to which he noted, ``Both the papers by de~Sitter [1917b] and Eddington [1923] lack this assumption on the ``state of rest'' of stars -- by the way the only possible one compatible with the homogeneity of space and time. Without such an assumption nothing can be known about the redshift, of course.'' He then went on to determine the redshift for this system, noting afterwards, ``All stars of our system $\Sigma$ flee from any arbitrary star in radial directions; there is inherent in matter a universal tendency to expand which finds its expression in the `cosmological term' of Einstein's law of gravitation.'' 

These attempts by Weyl and Eddington, to explain the observed redshifts through de~Sitter's solution were published in the spring of 1923, a year \textit{after} Friedmann's first expanding universe solution. The previous September, \cite{Einstein1922} submitted his first remarks on Friedmann's paper, which began, ``The results contained in the cited work relating to a non-stationary world seem suspicious to me. In fact, it is evident that the given solutions are incompatible with the field equations\ldots'' and he went on to state his reason. As we know \citep{Ellis1999}, Friedmann then arranged for an explanation to reach Einstein, who, in the following May, retracted his initial criticism, noting that his objection was based on a miscalculation, of which he had been convinced through Friedmann's letter. He corrected himself as follows \citep{Einstein1931b}: ``I consider Mr. Friedmann's results correct and clarifying. It is shown that, in addition to the statical, the field equations admit dynamical (therefore, varying in the time-coordinate) centrally-symmetric solutions for the spatial structure.''

It's noteworthy, that here again Einstein did not balk when describing \textit{the} spatial structure as dynamical, even to the point that he added the parenthetical note specifying that by this he meant specifically that it varies with ``the'' time-coordinate. Therefore, the solution's Lorentz covariance notwithstanding, when it came to cosmology (which is the only place it matters, since relativistic space-time is Lorentzian, so any frame of reference can be used to coordinate events locally) Einstein had once more fully supported the assumption of a cosmic---i.e. absolute---time.

And in this, it seems he was taken with Friedmann's solution, as the alternative to be supported if the spatial structure proved to be dynamical. In particular, there is evidence that Einstein already had in mind to support the oscillatory form of the solution which \citet{Friedmann1922} discussed at the end of his paper:
\begin{quote}
It is left to remark that the ``cosmological'' quantity $\lambda$ remains undetermined in our formulae, since it is an extra constant in the problem; possibly electrodynamical considerations can lead to its evaluation. If we set $\lambda=0$ and $M=5.10^{21}$ solar masses, then the world period becomes of the order of 10 billion years. But these figures can surely only serve as an illustration for our calculations.
\end{quote}

The evidence suggesting Einstein had this particular solution in mind as the preferred dynamical alternative to his statical model, and that he didn't waver in his choice for the eight-year period from 1923 to 1931, is as follows. First of all, he published nothing directly on the cosmological problem between his second note on Friedmann's solution \citep{Einstein1923b}, which arrived at \textit{Zeitschrift f\"{u}r Physik} on May 31, 1923, and his famous retraction of the cosmological constant following Hubble's discovery, in which he added his support to this solution of Friedmann's, with positive curvature and $\lambda=0$ \citep{Einstein1931a}. The second bit of evidence is contained in a postcard he sent to Weyl on May 23, 1923---\textit{just eight days} before the note on Friedmann's paper arrived at \textit{Zeitschrift f\"{u}r Physik}. There, \citet{Einstein1923a} discussed field theory, noting, e.g., that ``perhaps the field theory has already given away everything that lies in its possibilities'', and that he now believes ``all these attempts towards a purely formal basis will not lead to further physical knowledge.'' He then concluded with a typically cryptic note on cosmology, which indeed wouldn't have made any sense to anyone who hadn't read the last section of \citeapos{Friedmann1922} paper: 
\begin{quote}
With reference to the cosmological problem, I am not of your opinion. Following de~Sitter, we know that two sufficiently separate material points are accelerated from one another. If there is no quasi-static world, then away with the cosmological term.
\end{quote}

To Weyl, who was used to thinking of the $\lambda$-term as describing a ``universal tendency to expand'', this final remark must have seemed self-contradictory. Having no knowledge of Friedmann's solution, and having been immersed in the development of a theory that would describe systematic redshifts \textit{in an expanding universe that fundamentally} relied \textit{on the cosmological constant to drive expansion}, Weyl was confused. True:---geodesics would be accelerated from one another in de~Sitter's world; but should this be considered a mark against it? Weyl's answer was: Not if the redshifts should be taken to indicate that the Universe actually \textit{is} expanding. 

The following year, \citet{Weyl1924} published a dialogue, ``Inertial Mass and Cosmos'', staged between saints Peter and Paul, proving with poetic grace that he had no idea what Einstein meant when he rejected the cosmological term, as he indeed had no knowledge of \citeapos{Friedmann1922} solution. The dialogue opens with a discussion of Mach's principle and inertia, in which the ``apostate'' Paulus, who presents Weyl's ``heretical'' views against those of the ``Relativity Church'', is clearly setting up an argument that the phenomenon of acceleration in de~Sitter's world is something perfectly natural, due to the general relativistic ``guiding field.'' When the discussion moves on to cosmology, Paulus describes Weyl's ``de~Sitter cosmology,'' in which a bundle of geodesics, cut out from half the de~Sitter hyperboloid,\footnote{de~Sitter space is a one-sheeted hyperboloid in Minkowski space.}  extends down into the infinite past, sharing a common asymptote, and that they spread upwards from there like a fan over the whole hyperboloid, noting in particular that this bundle describes ``the normal undisturbed state of each star in the system.''
To this, Petrus objects in a manner very reminiscent of the statement in \citeapos{Einstein1923a} postcard: ``If the cosmological term does not help lead through to Mach's principle, then I consider it completely useless and I'm for the return to the elementary cosmology''---where, by ``elementary cosmology,'' he means a special relativistic world. The paper discusses only the three cosmological models: the ``elementary cosmology,'' Weyl's ``de~Sitter cosmology,'' and Einstein's statical model. From this, it's clear that when Einstein rejected the cosmological constant in his letter, Weyl understood him to mean that if his static, spherical model had to be abandoned, he would sooner support a special relativistic cosmology than de~Sitter's solution---in which the $\lambda$-term clearly didn't do its intended job---and Weyl had no knowledge of Friedmann's solution.
Naturally, Paulus objected that the return to special relativity would be too hasty. He raised a number of points, but the main one was the measured redshifts of the spiral nebulae. He argued that de~Sitter's solution is the only one of the three hypotheses that had been put forward as a description of the world structure, that could naturally and sufficiently account for it, finally concluding,--- 
\begin{quote}
If I think about how, on the de~Sitter hyperboloid the world lines of a star system with a common asymptote rise up from the infinite past, then I would like to say: the World is born from the eternal repose of ``Father \"{A}ther''; but once disturbed by the ``Spirit of Unrest'' (\textit{H\"{o}lderlin}), which is at home in the Agent of Matter, ``in the breast of the Earth and Man'', it will never come again to rest.
\end{quote}

Whether Einstein had a reaction to this paper, or remained as unaware of it as Weyl was of Friedmann's work, is unknown. As we've noted, he waited to announce his support of Friedmann's solution until after Hubble's empirical results were published. Whether this was out of caution, as he'd rather not abandon the cosmological term and his own cosmological model to inconclusive evidence, or he was by then too busy with other concerns, are things we can only speculate on. But while Einstein did move on to work in areas such as quantum statistics and unified field theory, other important work on the cosmological problem was taking place, and it's worth briefly recounting this.

First of all, in the same year that Weyl's dialogue was published, \citet{Friedmann1924} discovered his second solution, which then went completely unnoticed.

\citet{Lemaitre1925} soon discovered the line-element (coordinate system) that's relevant in the description of Weyl's ``de~Sitter cosmology''---i.e. the description of exponentially expanding Euclidean space that was eventually used most prominently in the steady state theory advanced by \citet{Hoyle1948} and \citet{Bondi1948}. Indeed, this solution cuts a bundle of geodesics from half the hyperboloid roughly as Weyl had specified. In the paper's conclusion, Lema\^{i}tre discussed two points: ``first the field is not static and secondly, the space has no curvature.'' The first point, he noted, could probably be accepted, quoting \citet{Eddington1923} as we have above, from the passage where he suggests that the non-statical nature of de~Sitter's world could be considered an advantage, and then adding, ``Our treatment evidences this non-statical character of de~Sitter's world which gives a possible interpretation of the mean receding motion of spiral nebulae.'' However, he stresses: ``The second point, on the contrary, seems completely inadmissible. We are led back to the euclidean space and to the impossibility of filling up an infinite space with matter which cannot but be finite. de~Sitter's solution has to be abandoned, not because it is non-static, but because it does not give a finite space without introducing an impossible boundary.''

In the history of cosmology, this paper is really only important because the work must have led Lema\^{i}tre to his rediscovery of \citeapos{Friedmann1922} original solution \citep{Lemaitre1927}, which played a central role in the development of cosmology after Hubble confirmed the redshift-distance relation; for, neither paper by \citet{Lemaitre1925,Lemaitre1927} was really read by anyone else at the time.\footnote{In fact, although these papers remained generally obscure, it is interesting to note that he did manage to solicit Einstein to read his 1927 paper. However, when Lema\^{i}tre sought him out at the Fifth Solvay Conference in Brussels in 1927 to ask what he thought, Einstein is famously quoted as having said, ``Your calculations are correct, but your physical insight is abominable'' (e.g. \citealt{Smith1990}). 

While this statement is usually given as evidence that Einstein, in 1927, still refused to consider the possibility that the Universe is expanding \citep{Smith1990,Kragh2003,Farrell2005,Bartusiak2009}, the point would contradict what we've inferred above, from \citeapos{Einstein1923b} postcard to Weyl, as well as his second note on \citeapos{Friedmann1922} paper, which not only called Friedmann's result ``correct'', but actually ``clarifying'' \citep{Einstein1923a}. Einstein even told Lema\^{i}tre about Friedmann's paper during this meeting in Brussels \citep{Kragh2003,Farrell2005,Bartusiak2009}, so it isn't that he had simply forgotten about it in the meantime. Therefore, from what we know of Einstein's opinions on cosmology and the cosmological constant, the most plausible reason for why he would have considered Friedmann's results ``clarifying'', but Lema\^{i}tre's physical insight ``abominable'', while acknowledging that the equivalent mathematics of the two were ``correct'', seems to be that Lema\^{i}tre had failed to recognise that the cosmological constant could be set to zero. 

The specific model described by \citet{Lemaitre1927} expanded from an Einstein universe in the infinite past, to a de~Sitter universe in the infinite future. From Einstein's standpoint, the $\lambda$-term had been justified only so long as the evidence supported a quasi-static world: if the Universe were expanding, then the only acceptable solution was \citeapos{Friedmann1922} with $\lambda=0$. \citet{Lemaitre1927} completely failed to recognise this when he described a universe that would fundamentally expand \textit{because of} a repulsion due to the cosmological constant. Einstein had already discussed this with Weyl, and had made up his mind that ``if there is no quasi-static world, then away with the cosmological term''. Thus, it should be no surprise that Einstein would have found Lema\^{i}tre's physical insight ``abominable''---and that this would have had nothing to do with a refusal on Einstein's part to admit that the Universe could be expanding, or an ignorance of the evidence that had motivated Weyl and Eddington to consider such an interpretation of the redshifts already in 1923. Even so, Lema\^{i}tre came away from the encounter with an impression that Einstein's negative opinion of the paper was a result of his being ``not current with the astronomical facts'' \citep{Bartusiak2009}.} We'll discuss the discovery of \citeapos{Lemaitre1927} paper in more detail in \S\,\ref{sec:4.1} below.

Probably the next most significant advance came in 1928, when Robertson independently discovered \citeapos{Lemaitre1925} line-element for de~Sitter space. Robertson, who was unaware of Lema\^{i}tre's earlier work,\footnote{Robertson discovered \citeapos{Lemaitre1925} paper only the following year. In 1929, he wrote a remarkably complete paper \citep{Robertson1929}: he derived the general line-element for spatially homogeneous and isotropic universes---i.e. the general line-element of Friedmann-Lema\^{i}tre-Robertson-Walker cosmological models---discussing those with and without the additional assumption of stationarity that was later required by the ``perfect'' cosmological principle of the steady state theory \citep{Hoyle1948,Bondi1948} \textit{and} he managed to discover nearly all the relevant references (excepting \citeapos{Lemaitre1927} paper); e.g., he noted, with apologies for having omitted the citation in \citep{Robertson1928}, that \citet{Lemaitre1925} had discovered these coordinates, and even cited \textit{both} papers by \citeapos{Friedmann1922,Friedmann1924}, the first of which would become generally known only the following spring, while the latter would remain generally unknown well into the 1930s. We'll discuss this last point further below, as its importance in the history of Einstein's cosmological considerations is significant.} wasn't troubled by it in the same way that Lema\^{i}tre had been; for, he noted, it may be objected that ``the world is on this interpretation of infinite extent, but, as will be seen later, its closed character is maintained in the sense that the only events of which we can be aware must occur within a sphere of finite radius.'' He therefore moved on to discuss the properties of this model. 

Most significantly, \textit{he fully anticipated Hubble's Law}: in a section on the Doppler effect, he determined a first-order linear relationship between velocity and distance in this solution, stating \citep{Robertson1928}:
\begin{quotation}
If we assume that there is no systematic correlation of coordinate velocity with distance from the origin [no systematic proper radial velocity], we should expect that the Doppler effect would indicate a residual positive radial velocity of distant objects\ldots Although we cannot, from Doppler effect alone, distinguish between proper motion and this distance effect, we should nevertheless expect a correlation
\[v\simeq c\,{l \over \mathrm{R}}\]
between assigned velocity $v$, distance $l$, and the radius of the observable world R.
\end{quotation}
At this point, he noted that \citet{Weyl1923b} ``deduced a similar relation on the de~Sitter hypothesis, which is also given by [this velocity-distance relation] to first order, assuming that the geodesics concerned form a diverging pencil of geodesics.'' \citet{Weyl1930} later showed that \citeapos{Robertson1928} redshift derivation was equivalent to his.

It stands to reason that \citet{Hubble1929} was aware of this theoretical work, because although all he bothered to cite were a handful of astronomy papers, he concluded his famous announcement by noting,
\begin{quotation}
The outstanding feature, however, is the possibility that the velocity-distance relation may represent the de~Sitter effect, and hence that numerical data may be introduced into discussions of the general curvature of space. In the de~Sitter cosmology, displacements of the spectra arise from two sources, an apparent slowing down of atomic vibrations and a general tendency of material particles to scatter. The latter involves an acceleration and hence introduces the element of time. The relative importance of these two effects should determine the form of the relation between distances and observed velocities; and in this connection it may be emphasized that the linear relation found in the present discussion is a first approximation representing a restricted range in distance.
\end{quotation}
We've discussed the theory behind this entire paragraph in detail. Hubble's discovery, while it is one of the most important empirical results in history, shouldn't have surprised anyone---\textit{and it likely didn't}. The theorists who were working on the cosmological problem of relativity theory throughout the 1920s had anticipated it---and the theoretical possibility of expansion, together with its empirical support, was publicised in both \citeapos{Eddington1923} \textit{Mathematical Theory of Relativity} and \citeapos{Weyl1923a} \textit{Raum, Zeit, Materie} (in the fifth edition)---so anyone who studied relativistic cosmology in either English or German in the 1920s should certainly have known to expect Hubble's result. The importance of Hubble's ``discovery'' was therefore rather that it finally provided the conclusive empirical evidence of  something that everybody already knew.

\section{The shift}
\label{sec:4}
The most interesting points discussed in the previous section are, that already in 1923 Einstein had made his decision to support \citeapos{Friedmann1922} solution with $\lambda=0$, should it turn out that the Universe isn't static, and that the theorists working on the cosmological problem of relativity theory throughout the 1920s were open to the possibility that the Universe is expanding, and had therefore anticipated Hubble's result. 

Nevertheless, we know that there was a great shift in thinking following Hubble's discovery---and not only amongst those who \textit{hadn't} been working on the problem, for, even those who'd known about the theoretical work of the 1920s didn't smoothly transition as a result of the empirical confirmation: within a short time, no one seriously considered de~Sitter's vacuum solution to be a legitimate model for our Universe until it was resurrected nearly twenty years later by the steady state theorists \citep{Hoyle1948,Bondi1948}, as the only stationary expanding universe model in agreement with their ``perfect cosmological principle''.

The shift in thinking wasn't due to a change of heart---e.g., many still thought of the cosmological constant as the fundamental driving force of cosmic expansion; and in any case everyone did \textit{continue on} thinking that the Universe expands---rather, the shift occurred as a result of the discovery of \citeapos{Friedmann1922} and \citeapos{Lemaitre1927} solutions, which had lain dormant until they were brought to general awareness in the spring of 1930---more than a year after Hubble's publication.

\subsection{The gun was loaded\ldots}
\label{sec:4.1}
The discovery of \citeapos{Friedmann1922} and \citeapos{Lemaitre1927} papers, and the shift in thinking that took place as a result, was recorded in the papers written at the time. But in order to get the story straight, it's important to take a brief step back, in order to see where Eddington and de~Sitter had already been headed in their theoretical investigations. Unlike Weyl, Eddington was actually unhappy with the fact that de~Sitter's world is empty, and had already expressed his hope of finding an intermediary solution \citep{Eddington1923}: 
\begin{quote}
It seems natural to regard de~Sitter's and Einstein's forms as two limiting cases, the circumstances of the actual world being intermediate between them. de~Sitter's empty world is obviously intended only as a limiting case; and the presence of stars and nebulae must modify it, if only slightly, in the direction of Einstein's solution. Einstein's world containing masses far exceeding anything imagined by astronomers, might be regarded as the other extreme---a world containing as much matter as it can hold.
\end{quote}
And this is precisely what Eddington was working on when, in the spring of 1930, Lema\^{i}tre sent him (for a second time) a reprint of his 1927 paper, which provided the complete description of a universe that would expand from an Einstein universe in the infinite past, to a de~Sitter universe in the infinite future.

There has been some question about whether Eddington actually read \citeapos{Lemaitre1927} paper \textit{in} 1927, but failed to recognise its importance and then simply forgot about it until Lema\^{i}tre reminded him of it in a letter he sent in the spring of 1930. The confusion over this point is due to later recollections of George McVittie. In an obituary written for Lema\^{i}tre, \citet{McVittie1967} recalled ``the day when Eddington, rather shamefacedly, showed me a letter from Lemaitre which reminded Eddington of the solution to the problem which Lemaitre had already given. Eddington confessed that, though he had seen Lemaitre's paper in 1927, he had completely forgotten about it until that moment.'' While this statement tends to convey a sense that Eddington might actually have read the paper when he saw it in 1927, when McVittie was later questioned,\footnote{Interview of Dr.\ George C.\ McVittie by David DeVorkin on March 21, 1978, Niels Bohr Library \& Archives, American Institute of Physics, College Park, MD USA, \url{http://www.aip.org/history/ohilist/4774.html}} his response was less certain: ``Lemaitre says in a letter to Eddington that he sent Eddington a reprint [when the paper came out in 1927], and I think Eddington had simply forgotten. In fact, I think Eddington told me that. I think Eddington's words were, as far as I recall them, `I'm sure Lemaitre must have sent me a reprint, he's just sent me another, but I'd forgotten all about it.'{''} 

Since \citet{Eddington1923} had anticipated \citeapos{Lemaitre1927} solution, already considering the redshift data as evidence of cosmic expansion, and in light of his reaction when he did read the paper in 1930, it seems unlikely that he would have read it in 1927. Eddington's reaction when he finally read the paper is actually noted at the beginning of a paper in which he examined the instability of Einstein's world \citep{Eddington1930}; but there is yet more detail in the minutes from the May 9 meeting of the Royal Astronomical Society (RAS), where the paper was discussed \citep{RAS1930}. There, Eddington stated,
\begin{quotation}
Some time ago I conjectured that Einstein's spherical world might be unstable. More recently I thought I saw a way to settle the question mathematically. I was working on this problem with Mr. McVittie, and we had nearly reached the solution when I learnt of a remarkable paper by Abbe G. Lema\^{i}tre, of Louvain, published in 1927, which contained all the necessary mathematics. He does not say explicitly that Einstein's world is unstable, but it follows immediately from his equations. I think this makes a great difference to our outlook; Einstein's solution gave the only possible condition of equilibrium of the universe, and now this proves to be unstable. de~Sitter's is also reckoned technically as an equilibrium solution, but it is a bit of a fraud; being entirely empty, there is nothing in his world whose equilibrium could possibly be upset. In saying this I am not disparaging it, because it is much more interesting than a genuine equilibrium solution would have been.

To discuss stability we must have a range of solutions, and Lema\^{i}tre's work provides this. He treats of a world whose radius is a function of time. Instead of having to choose between Einstein's and de~Sitter's worlds our conclusion now is that the universe started as an Einstein world, being unstable it began to expand, and it is now progressing towards de~Sitter's form as an ultimate limit.
\end{quotation}
He went on to discuss some more details of the model, which aren't relevant for our purposes---but de~Sitter's comment in reaction to this statement is of interest:
\begin{quotation}
I agree entirely with Prof.\ Eddington's remarks. Einstein's solution gives a world full of matter, but no motion; mine gives a world full of motion, but no matter. I had been pursuing the same kind of investigation as has Prof.\ Eddington when he sent me Lema\^{i}tre's paper, and have found the same as Eddington.
\end{quotation}
And so we'll leave off with Eddington, who probably neglected to read \citeapos{Lemaitre1927} paper when he first received a copy, and move to de~Sitter, where we'll find a number of interesting points---but before we do that, one additional statement will suffice to show that it was indeed Lema\^{i}tre's discovery, rather than Hubble's, that served as the turning point \citep[p.~60]{Eddington1933}:
\begin{quotation}
The Einstein configuration was the one escape from an expanding or contracting universe; by proving it to be unstable, we show that it is no more than a temporary escape\ldots by its own resources [physical theory] has been guided into the road to a non-static universe. Realising that some degree of expansion (or contraction) is inevitable, we are much more inclined to admit the recession of the spiral nebulae as an indication of its magnitude.
\end{quotation}
So it was, by its own faculties, that relativistic cosmology was seen to lead inevitably to the prediction of an expanding universe.

Ten days after the May 9 RAS meeting, \citet{deSitter1930} communicated a paper ``On the distances and radial velocities of extra-galactic nebulae, and the explanation of the latter by the relativity theory of inertia'', to the \textit{Proceedings of the National Academy of Sciences of the United States of America} (\textit{PNAS}). In the first four sections he gave a state-of-the-art discussion of the determinations of the redshift-distance relation by ``Hubble, Lundmark, Shapley and others'', before turning his attention to a theoretical investigation. In the fifth section, he began discussing the solutions to the Einstein field equations that have complete spherical symmetry in space, and focussed on the two statical solutions of this form---Einstein's, and his own---noting that these are the only two solutions to Einstein's equations that can be written with complete spherical symmetry in space of finite radius, with a line-element that depends only on the three spatial coordinates (i.e., not on time). He then showed that both of these solutions must be rejected on empirical grounds;---Einstein's, because there is no systematic velocity-distance relation, and his own empty universe, because although there is an approximately linear velocity-distance relation, the observational evidence showed that the world is in fact practically full of matter, as in Einstein's solution.

At this point, de~Sitter finally noted that a dynamical solution of the field equations in which spherically symmetric space is filled with matter, was already ``given by Dr. G. Lema\^{i}tre in a paper published in 1927, which had unfortunately escaped my notice until my attention was called to it by Professor Eddington a few weeks ago.'' In an endnote, de~Sitter cited \citeapos{Lemaitre1927} paper, and noted that \citet{Friedmann1922} had previously found a ``similar solution''.

In the remainder of the paper, de~Sitter went on to analyse the astronomical data in accordance with Lema\^{i}tre's solution. The only other point he made that's notable in our discussion is a remark at the end: ``The constant $\lambda$, which is a measure of the inherent expanding force of the universe, is still very mysterious, and it is difficult to see what its real meaning is. It might even be thought to be one constant too many, unless we may hope that it will ultimately be found to be in some way connected with Planck's constant $h$.'' Therefore, while de~Sitter's sceptical opinion of the $\lambda$-term had evidently lingered, it had relaxed by some degree, and he did recognise that the cosmological constant provides an ``inherent expanding force''.

And so it was, that the works of both \citet{Friedmann1922} and \citet{Lemaitre1927} came to be known by those who would influence the further development of theoretical cosmology. In hindsight, however, there remains a peculiarity: the papers by \citet{Friedmann1922}, \citet{Lemaitre1927}, \citet{Eddington1930}, and \citet{deSitter1930} all assume that space should be spherically closed. On the one hand, given \citeapos{Einstein1917} earlier cosmological considerations, it really isn't surprising that the emphasis on solutions in which space is finite remained as part of the early expanding universe models. After all, it may not have been obvious that dynamical solutions with open space shouldn't suffer the same problems that Einstein had found when he considered the statical case. But, on the other hand, \citet{Friedmann1924} had already discovered an acceptable solution with open space; and \citeapos{Lemaitre1925} reaction to the line-element he'd discovered for de~Sitter space-time, describing Euclidean space that expands exponentially in the course of cosmic time, certainly confirms this suspicion. The key to this particular puzzle is to note that \textit{\citeapos{Friedmann1924} second solution remained widely unknown}. And indeed, it would be difficult to properly understand Einstein's next step on the cosmological problem without knowing that \textit{he was unaware of Friedmann's second paper}; in fact, we've found no evidence that he ever knew of it.

\subsection{\ldots Einstein pulled the trigger\ldots}
\label{sec:4.2}
We'll show in due course that \citeapos{Friedmann1924} second paper was unknown to Einstein, de~Sitter, and others during the early 1930s. For now, in order to make sense of his cosmological considerations at the time, and to be able to properly discuss the influence Einstein had on the paradigm shift that eventually took place in cosmology, it's enough to know that, going forward, the closed expanding universe model was all he knew about. 

Now, in the years separating Einstein's first \citep{Einstein1917} and second \citep{Einstein1931a} papers on cosmology, we've seen that the cosmological term he originally introduced, had come to be understood as describing a repulsive force that would \textit{drive} cosmic expansion. For example, while \citet{deSitter1930} later described it as ``a measure of the inherent expanding force of the universe'', one of his main criticisms of Einstein's paper in 1917 was the seemingly arbitrary introduction of the $\lambda$-term, which he said ``detracts from the symmetry and elegance of Einstein's original theory, one of whose chief attractions was that it explained so much without introducing any new hypothesis or empirical constant'' \citep{deSitter1917a}.

It turns out that Einstein agreed with the earlier point, but not the later one. Indeed, as discussed above, he had already decided, by May 1923, and much to Weyl's confusion, that ``if there is no quasi-static world, then away with the cosmological term.''	

In 1931, following Hubble's empirical confirmation of the expansion of space---and in fact following a visit to Mount Wilson Observatory during his winter stay in Pasadena---he finally publicised this opinion. \citeapos{Einstein1931a} paper is really unremarkable apart from the fact that he rejected the $\lambda$-term at a time when everyone else was considering general solutions with cosmic expansion fundamentally \textit{driven} by $\lambda$. The paper was essentially a restatement of \citeapos{Friedmann1922} solution with $\lambda=0$, taken to be justified as a realistic model by \citeapos{Hubble1929} empirical result. The only significant point he made was that Hubble's result indicates that the universe is indeed \textit{not} static, and that this can be taken, together with Friedmann's solution, to show that the cosmological constant is not necessary after all. 

\citet{Einstein1931a} mentioned the $\lambda$-term twice in the paper, as follows: after noting \citeapos{Friedmann1922} and \citeapos{Hubble1929} results, he first remarked, ``With these understandings, one must ask oneself whether the facts can be accounted for without the introduction of the anyhow theoretically unsatisfactory $\lambda$-term''; then, after showing that this can indeed be done, in the paper's final sentence he concluded that ``Above all, it is remarkable that Hubble's new facts allow general relativity theory to seem less contrived (namely, without the $\lambda$-term), as the postulate of the quasi-static nature of space has moved into the distance.''

Reading these lines, one almost has to wonder whether he was even \textit{aware} that his proposal directly opposed the prevailing conception of cosmic expansion; for in contrast to 1917, he offered little speculation at all in the three-page paper, and certainly made no argument \textit{against} the common idea of cosmic repulsion, which he left unmentioned. Indeed, he gave no reason at all why he thought the $\lambda$-term was ``theoretically unsatisfactory'', but merely expressed his distaste for the term adjectivally, i.e. as a matter of fact.

Fortunately, a letter he sent to Tolman on June 27, 1931 has survived, which provides some clarification. There, he began by stating that his paper ``contains only the point that the $\lambda$-term is unnecessary if one allows for solutions with time-variable world radius'', which, he added, is ``really incomparably more satisfying.'' This is helpful to know, because, as noted above, the paper certainly \textit{appears} unremarkable otherwise; so, `from the horses mouth' so to speak, we have that assessment confirmed. Einstein went on in this letter to discuss the problem of the singularity at the starting point in the model, conjecturing that the mathematical idealisation is useless because of an unevenness in the density of matter; therefore, in the neighbourhood of that point in time, space would have to have been relatively very small and inhomogeneous.\footnote{\citet{Hawking1967} eventually disproved this conjecture.}  He then noted a suggestion by Lindemann, that planet formation could have to do with these conditions. 

It was only in a postscript that he finally clarified his intention to reject ``the somewhat adventurous way of interpreting Hubble's line shifts,'' which, he noted, was something that should be apparent. Therefore, while his paper didn't mention it, \citet{Einstein1931a} was indeed aware of the interpretation of cosmic expansion he was rejecting along with the cosmological constant.
	
\citeapos{Einstein1931a} intention in supporting the particular model that he did has always been a source of confusion due to the particular nature of that solution. It describes space as expanding from a singularity for only a finite time before coming to a halt, and then contracting back on itself in time-symmetric fashion. There's a very obvious possibility that the universe could rebound (if the singularity could be avoided) and go through a series of expansions, contractions, and ``big bounces''---and Einstein surely had to be aware of this, as \citet{Friedmann1922} had actually described the closed model with $\lambda=0$ as ``periodic'' in his original paper. But there is no indication, either in the paper \citep{Einstein1931a} or in his letter to Tolman \citep{Einstein1931b}, that Einstein cared one iota for the possibility of periodic expansion, contraction, and rebirth. As he stated in his letter to Tolman, his sole purpose was to make the point that ``the $\lambda$-term is unnecessary if one allows for solutions with time-variable world radius''---a point he made using the only such solution he knew of at the time.

Yet for a while, the ``oscillating'', ``quasi-periodic'' model favoured by Einstein was discussed. de~Sitter, who nearly rivaled Dumas in his ability to churn out pages at this time, soon published ``Some further computations regarding non-static universes'', as he had neglected to pay particular attention to the $\lambda=0$ oscillatory case in previous work \citep{deSitter1931}. The paper begins:
\begin{quotation}
The non-static solution of the field equations of the general theory of relativity, of which the line-element is
\[sd^2=-R^2d\sigma^2+c^2dt^2,\]
$R$ being a function of $t$ alone, and $d\sigma^2$ being the line-element of a three-dimensional space of constant positive curvature with unit radius,\footnote{It's noteworthy that, as of August 7, 1931, de~Sitter hadn't yet begun to discuss solutions with non-positive curvature.}  have been investigated by Friedmann in 1922 and independently by Lema\^{i}tre in 1927, and have attracted general attention during the last year or so. Einstein has lately expressed his preference for the particular solution of this kind corresponding to the value $\lambda=0$ of the ``cosmological constant''. This solution belongs to a family of oscillating solutions, which were not included in my discussion in \textit{B.\ A.\ N.}\ 193.
\end{quotation}
Thus, de~Sitter immediately referred to Einstein's preferred model as ``oscillating'', even though \citet{Einstein1931a} had expressed no interest in this aspect---in fact, he didn't even mention the contracting phase in his letter to Tolman \citep{Einstein1931b}---and we know that his reason for this preference was simply that he wanted to set $\lambda=0$.

Tolman's reaction to the possibility of an oscillating universe was more enthusiastic yet; for while \citet{Einstein1931b} had made no mention of this in his letter, but had expressed his `incomparable satisfaction' at finally being able to reject the cosmological constant, \citeapos{Tolman1931a} response, on September 14, 1931, began, ``When I first saw \textit{your proposed quasi-periodic solution} [emphasis added] for the cosmological line element, I was very much troubled by the difficulties connected with the behaviour of the model in the neighborhood of the points of zero proper volume.'' He went on to say that through Einstein's remarks about inhomogeneity, and Lindemann's suggestion that the early state could have something to do with the formation of the planets, he felt more comfortable about the difficulty. In fact, in addition to Einstein's remarks, he added that ``from a physical point of view contraction to a very small volume could only be followed by renewed expansion'', noting that he had just sent an article to the \textit{Physical Review} discussing just such a possibility.

Indeed, in the last sentence of the introduction to \citeapos{Tolman1931b} ``On the theoretical requirements for a periodic behaviour of the universe'', he cited Einstein as his motivation: ``Recently\ldots a simple model of the universe has been discussed by Einstein which exhibits a possibility for quasi-periodic solutions of a type which must now also be discussed'', referencing both \citeapos{Einstein1931a} and \citeapos{deSitter1931} papers. 

Further down in his letter, \citet{Tolman1931a} remarked on some of the details of his paper; but before getting to that, he commented on Einstein's rejection of the cosmological constant. He listed a few points in favour of this, mostly having to do with parsimony---``the fundamental equations of the theory are simplified, the conclusions drawn from them are rendered less indeterminate, and it becomes no longer necessary to inquire into the significance and magnitude of what would otherwise be a new constant of nature''---but stated as well what he considered ``one fairly strong'' point against it: 
\begin{quote}
since the introduction of the $\lambda$-term provides the most general possible expression of the second order which would have the right properties for the energy-momentum tensor, a definite assignment of the value $\lambda=0$, in the absence of an experimental determination of its magnitude, seems arbitrary and not necessarily correct. 
\end{quote}

Much as in the earlier episode with Weyl, Einstein's intention seems to have been misunderstood. He had expressed no interest in quasi-periodicity, yet Tolman seems to have thought it was the oscillatory nature of the model that interested him---and indeed that is the aspect of Einstein's preferred model that Tolman was the most enthusiastic about. But Einstein stated overtly, that the only point of his paper was to show that if solutions with time-variable radius are to be considered, then the $\lambda$-term is unnecessary---``if there is no quasi-static world, then away with the cosmological constant''!---and that ``Hubble's line shifts'' could be interpreted in a way that he considered less adventurous. This response must therefore have been deflating: in Tolman's opinion, the field equations of general relativity were originally incomplete without the $\lambda$-term, and setting the cosmological constant equal to zero without having any empirical evidence to do so, would be an arbitrary and potentially false move.---And indeed, the cosmological evidence now indicates that it \textit{is} wrong to set $\lambda=0$. This point will be discussed later on, when we've completed our discussion of Einstein's rejection of the cosmological term, and the paradigm shift that came with it.

So far in this discussion, we've noted some confusion over Einstein's decision to support the particular Friedmann model he did in 1931, and argued that his reason for this had nothing to do with an interest in periodic oscillations, as he only wished to reject the $\lambda$-term. The argument has so far been based on some remarks to that effect in a letter he wrote to Tolman, as well as the content of the paper in which he made the point. In that paper, the reason why he'd want to reject the $\lambda$-term is certainly made obvious enough---he considered it theoretically unsatisfactory and thought its introduction had made the general theory of relativity seem contrived---but he offered no reason whatsoever for why he thought this. We can therefore only speculate based on what he wrote elsewhere, but it seems reasonable to suggest that his main concern was to describe a cosmological model in which the world-structure was determined by a finite density of matter, which had been sufficient cause to introduce the term \textit{just in case} the evidence should indicate a static cosmos. Indeed, the requirement of a finite density, and that ``the field should be due to the matter, without which it cannot exist'' \citep{deSitter1917a}, was the original reason why he felt the term's introduction had been justified, and any later remarks he made were always to that effect \citep[p.~685]{Einstein1932,Einstein1945,Einstein1949}. Furthermore, it's notable that \citeapos{Einstein1931a} two disparaging remarks against the term seem to have been taken directly from \citet{deSitter1917a}, from the very argument where he showed that \textit{just in case there is a positive cosmological constant} it's possible to derive a world model with ideal boundary conditions and complete relativity of time, which is \textit{perfectly devoid} of matter. In light of the fact that his \textit{main concern} seems to have been to describe cosmic structure as being fundamentally determined by a finite density of matter, this must have been, for Einstein, a strong point against the $\lambda$-term.

Therefore, it seems that Einstein's reason for \textit{wanting to reject} the $\lambda$-term was that de~Sitter's \textit{counter-example} had shown that his main concern was not met just in case---i.e. if, \textit{and only} if---it was introduced (which he originally considered sufficient reason to justify its introduction), so the term was theoretically unsatisfactory and, as de~Sitter had argued, made the theory seem contrived, i.e. `detracting from its original simplicity and elegance' \citep{deSitter1917b}. And for that reason, it really is no wonder that he would embrace the opportunity to support the $\lambda=0$ case of Friedmann's solution, which proved that his main concern could be met without the $\lambda$-term as long as the world-radius was time-variable. His postcard to Weyl \citep{Einstein1923b} shows that he was ready to do this already in 1923, and the fact that he waited until after he'd been to Mount Wilson and met Hubble shows just how strong his sense of pragmatism was, as he would not state his preference until the time-variable spatial structure it entailed had been confirmed by sufficiently reliable empirical evidence. 

This paints a picture of Einstein's thoughts on cosmology in 1931, that answers why he had wanted to reject the $\lambda$-term, and therefore why he chose to support the particular Friedmann model that he did. It also sets us up to understand why he thought the introduction of the $\lambda$-term was the biggest blunder he ever made in his life, which will become more obvious as we discuss other proposals that would make use of the term. However, this should already be somewhat evident from \citet{Tolman1931a} argument that, from a theoretical standpoint, he considered ``a definite assignment of the value $\lambda=0$, in the absence of an experimental determination of its magnitude, [to be] arbitrary and not necessarily correct.'' For, if we understand Einstein's opinion to be that $\lambda$ would have been justified if, and only if, it would support a finite matter density in a statical world---a point that immediately backfired in the worst possible way, through \citeapos{deSitter1917a} counter-example---then continued use of the term, for whatever reason, would have aggravated him, \textit{particularly} in light of the fact that it was now justified to achieve his main concern without the term.

In \S\,\ref{intro}, we noted that some have recently come to attribute Einstein's regret over having introduced the $\lambda$-term, to the fact that he might otherwise have predicted the expansion of space and gone down in history as the discoverer of the expanding universe \citep{Gribbin1986,Dainton2001,Penrose2005,Kolb2007}. We've now noted two strong points against this. First of all, it seems that Einstein put little stalk in pure speculation; that not only would his sense of verificationism have kept him from making such a proposal on purely theoretical grounds, but he would not even have seen the merit, from a physical point of view, in making such a prediction without good observational grounds to do so. Therefore, having decided already in 1923 that ``If there is no quasi-static world, then away with the cosmological term,'' he only officially announced his opinion after he had seen sufficiently convincing evidence that the world \textit{isn't} quasi-static. And secondly, as noted above (and we'll actually see more evidence of this below), the expansion of space \textit{was} seen by some as the inevitable outcome of having \textit{introduced} the $\lambda$-term, so the recent speculation has really got the situation completely backwards, since Einstein \textit{should} get credit, from that point of view, for predicting expansion.

This whole picture, however, and particularly our argument that Einstein was totally uninterested in the potential periodicity of the model, hinges on the observation that he was unaware of \citeapos{Friedmann1924} second paper, and the possibility that the spatial curvature could be zero or negative. Indeed, given that everyone today is aware of the general possibilities, not knowing the fact that in 1931 Einstein and his peers \textit{were} unaware of them should certainly lead one to think that the oscillatory nature of the specific model he preferred had been particularly appealing to Einstein---especially given that this \textit{was} the inference his peers made, being unaware of the extent of his fixation with rejecting the $\lambda$-term. \citet[pp.~176-178]{Steinhardt2007} indeed recently made just this point: ``He could have chosen Friedmann's ever-expanding universe, but he didn't. Instead, he fixed his attention on Friedmann's model of a closed, periodic (oscillatory) universe.''

But Einstein really didn't know of the other possibilities; and, as it turns out, when he did become aware of them, he immediately opened his support to the expanding models with general curvature---and $\lambda=0$. This did not take long to happen.

In July 1931, Heckmann published a paper ``On the metric of the expanding universe'' \citep{Heckmann1931}, which prompted \citeapos{Einstein1932} PNAS article---the two pages of which began, ``In a recent note in the \textit{G\"{o}ttinger Nachrichten}, Dr.\ O.\ Heckmann has pointed out that the non-static solutions of the field equations of the general theory of relativity with constant density do not necessarily imply a positive curvature of three-dimensional space, but that this curvature may also be negative or zero.'' In the next paragraph, they noted that while the data had indicated that the actual universe is non-statical, neither the sign nor the value of the curvature could be determined with the available data---so they proposed to represent those data with the assumption that there is no curvature; i.e., they proposed to describe three-dimensional expanding space as Euclidean. In the third paragraph, they noted that the cosmological constant was \textit{only introduced into the theory in order to account for a finite mean density of matter in a static universe}, and that in the dynamical case the same end could be reached without $\lambda$. And in the fourth paragraph, they wrote down the line-element for expanding flat space, along with the one relevant differential equation describing the evolution of the scale-factor that multiplies space, when pressure is neglected, from ``the field equations without $\lambda$". 

At this point, Einstein's rejection of the $\lambda$-term and the ``adventurous way to interpret Hubble's line shifts'' was complete. In the remainder of the paper, they went on to show, as they eventually concluded, that ``at the present time it is possible to represent the facts without assuming a curvature of three-dimensional space''---noting, however, that the curvature is determinable, and would be possible to determine in the future with more precise data.

Certainly, \citeapos{Einstein1932} reaction to \citeapos{Heckmann1931} paper indicates that neither had been aware of \citeapos{Friedmann1924} second cosmological solution; for if they'd known from Friedmann that the curvature could just as well be positive or negative, they surely could have deduced that it could be zero as well. And the fact that Einstein was so quick to abandon the ``oscillatory'' model when he learned that positive curvature wasn't necessary, agrees as well with what we've said about his motives for having favoured that model to begin with.

But the best proof that \citeapos{Friedmann1924} second paper was, and remained, unknown, is found in \citeapos{deSitter1933} ``Astronomical Aspect of the Theory of Relativity''. In the opening paragraph to the section on non-static solutions, he wrote down the details of the discoveries of the expanding models, eventually noting that ``Friedmann discusses the solutions of the field equations for different values of $\lambda$. Lema\^{i}tre considers only a positive $\lambda$. Both authors consider a positive curvature of space only. The fact that both $\lambda$ and the curvature may as well be negative or zero was only pointed out by Dr. Heckmann in July 1931.'' \footnote{Incidentally, in the preceding paragraph (the last one of the preceding section, in a note on the proof that \citeapos{Einstein1917} and \citeapos{deSitter1917a,deSitter1917b} solutions are the only possible static, homogeneous, and isotropic solutions with positive curvature), de~Sitter cited a paper by \citep{Robertson1929}---which \textit{did} already cite \citeapos{Friedmann1924} negative curvature solution. More than anything, this funny mix-up illustrates Robertson's uncommon appreciation amongst modern physicists, for knowing all the relevant literature, and for properly attributing priority to those who made the original discoveries in relativistic cosmology. This fact is also evident, e.g., in the detailed bibliography included in his unified account of ``Relativistic Cosmology'' in \textit{Reviews of Modern Physics} \citep{Robertson1933}. This is given as a list, broken up by year, of the major papers that contributed to the theory, with the full title and a brief remark on the main points discussed in each paper.} 

While Einstein, after 1932, never made such an explicit statement indicating that he remained unaware of \citeapos{Friedmann1924} second paper, we also don't know of any evidence that he ever \textit{did} know of it. For example, in the appendix to \textit{The Meaning of Relativity} on the cosmological problem, which he added in 1945, only \citeapos{Friedmann1922} first paper is cited. In any case, the second paper would have been far less meaningful to him, as the first solution was enough, together with Hubble's empirical confirmation of the expansion of space, to reject the cosmological constant. 

And while we know that cosmic expansion was eventually commonly conceived as being unrelated to the cosmological constant, and indeed that the $\lambda$-term was considered no more \textit{integral} to the field equations than, say, any higher-order term, the shift towards Einstein's way of thinking was still very gradual. Even so, it \textit{was} finally \textit{expected} that the assumptions of the Einstein-de~Sitter model would be confirmed by observations of type Ia supernovae when, in 1998, it was discovered that the cosmic expansion is \textit{actually} accelerating in a manner that fits with a positive cosmological constant \citep{Riess1998,Perlmutter1999}.

However, as we've seen, Tolman for one did not agree with Einstein's rejection of the cosmological constant. Eddington was actually incensed. And \citet{Chandrasekhar1983} recalled asking Lema\^{i}tre, sometime in the late fifties, what he judged as the most important change general relativity had wrought on our basic physical concepts,---to which the latter replied ``without a moments notice'' the cosmological constant. Indeed, the common rejection of $\lambda$'s possible fundamental importance came not with the bang that Einstein would have liked, but as a long drawn-out whisper.

\subsection{\ldots yet still the paradigm was slow to set.}
\label{sec:4.3}
We've just noted Lema\^{i}tre's remark to Chandrasekhar, that in the late 1950s he still thought of the cosmological constant as \textit{the} aspect of general relativity that had wrought the most important change to our basic physical concepts. And contrary to Einstein's opinion that the $\lambda$-term ``implies a considerable renunciation of the logical simplicity of the theory'' \citep[p.~685]{Einstein1949}, which he seems to have adopted from \citet{deSitter1917a}, Tolman's reaction to setting $\lambda=0$ in the field equations was that the choice is arbitrary and not justified by the facts. To Tolman, it didn't matter that Einstein had found a way to describe a finite density of matter in the universe without the $\lambda$-term: the field equations had been incomplete without it, and setting it equal to zero was a subjective choice with no scientific backing. 

In fact, as we now know, the evidence stands against Einstein's rejection. And in any case, arriving at the right idea for wrong reasons is not a good reason to reject that right idea. But Einstein used just this argument:---in the appendix on cosmology that he added to \textit{The Meaning of Relativity} in 1945, he argued that the $\lambda$-term seems so much less justified to introduce into the field equations ``since its introduction loses its sole original justification'' and would never have been introduced if Hubble's expansion had been discovered when general relativity was created \citep{Einstein1945}. \citeapos{Lemaitre1949a} response to this point of Einstein's was to note that ``The history of science provides many instances of discoveries which have been made for reasons which are no longer considered satisfactory. It may be that the discovery of the cosmological constant is such a case.''

Possibly the biggest critic of Einstein's decision to reject the cosmological constant was Eddington. He opposed Einstein's decision both at a fundamental level, and particularly when it came to interpreting cosmic expansion---which, as we've noted, he considered the inevitable outcome of physical theory \citep{Eddington1933}. For, Eddington had long conceived of the general relativistic field equations as an expression of the fact that length measurements are made always in relation to a finite constant directed radius of curvature. As he famously wrote in \textit{The Mathematical Theory of Relativity}, ``From this point of view it is inevitable that the constnat $\lambda$ cannot be zero; so that empty space has a finite radius of curvature relative to familiar standards. An electron could never decide how large it ought to be unless there existed some length independent of itself for it to compare itself with'' \citep{Eddington1923}. Even after Einstein rejected the term, Eddington continued arguing for this expression of the true meaning of the theory \citep{Eddington1933}:
\begin{quotation}
When once it is admitted that there exists everywhere a radius of curvature ready to serve as comparison standard, and that spatial distances are directly or indirectly expressed in terms of this standard, the law of gravitation $(G_{\mu\nu}=\lambda{g_{\mu\nu}})$ follows without further assumption; and accordingly the existence of the cosmical constant $\lambda$ with the corresponding force of the cosmical repulsion is established. Being in this way based on a fundamental necessity of physical space, the position of the cosmical constant seems to me impregnable; and if ever the theory of relativity falls into disrepute the cosmical constant will be the last stronghold to collapse. \textit{To drop the cosmical constant would knock the bottom out of space}.
\end{quotation}

It's worth noting in more detail Eddington's reasoning regarding the ``cosmical constant'', for in it lies a good point that appears particularly relevant in light of the current cosmological evidence. He noted \citep[p.~22]{Eddington1933} that Einstein's original law stated that $G_{\mu\nu}$ is always zero, so it would be independent of the other geometrical characteristic of empty space, $g_{\mu\nu}$,---whereas the amended law stated that the two are proportional. This relation, he added, places a significant limit on the geometrical properties allowed by nature. In contrast to Einstein's feeling (inherited from de~Sitter) that the $\lambda$-term was a complication to the logical simplicity of the original theory, Eddington considered the theory significantly tidied up with the introduction of the term. And what's more, he added,
\begin{quotation}
We have already said that the original term in the law [$G_{\mu\nu}$] gives rise to what is practically the Newtonian attraction between material objects. It is found similarly that the added term ($\lambda{g}_{\mu\nu}$) gives rise to a repulsion directly proportional to the distance. Distance from what? Distant from \textit{anywhere}; in particular distance from the observer\ldots Thus in straightening out his law of gravitation to satisfy certain ideal conditions, Einstein almost inadvertently added a repulsive scattering force to the Newtonian attraction of bodies. We call this force the \textit{cosmical repulsion}, for it depends on and is proportional to the cosmical constant\ldots In practical observation the farthest we have yet gone is 150 million light-years. Well within that distance we find that celestial objects are scatttering apart as if under a dispersive force. Provisionally we conclude that here cosmical repulsion has become dominant and is responsible for the dispersion.
\end{quotation}

It should be noted that the difference of opinion between Einstein and Eddington regarding the fundamental nature of Einstein's theory of gravitation was philosophical, rather than scientific---and each took the empirical evidence from cosmology to support his own philosophical stance on general relativity: Einstein found that with a time-variable spatial scale the $\lambda$-term was unnecessary and the theory could be stated more simply by rejecting it; whereas Eddington found that the theory was stated more simply \textit{with} the term, and noted the observed expansion confirmed the amended and simplified theory's prediction that there should be significant cosmical repulsion acting at large scales. He considered the observed expansion to be nearly as important a confirmation of Einstein's amended theory as the perihelion precession of Mercury's orbit:
\begin{quote}
It is well known that Einstein's law differs slightly from Newton's, giving for example an extra effect which has been detected in the orbit of the fast-moving planet Mercury; the cosmical repulsion is another point of difference between them, detectable only in the motions of remote objects. From a theoretical standpoint I think there is no more doubt about the cosmical repulsion than about the force which perturbs Mercury; but it does not admit of so decisive an observational test. As regards Mercury the theoretical prediction was quantitative; but relativity theory does not indicate any particular magnitude for the cosmical repulsion. A merely qualitative test is never very conclusive.
\end{quote}
In Eddington's opinion, the Nobel Prize-winning observation that the cosmic expansion \textit{is} accelerating at a rate consistent with a cosmological constant \citep{Riess1998,Perlmutter1999} would have been a great verification of the theory, and a demonstration of its predictive power.
We might say that, so long as the cosmical repulsion was confirmed only on a qualitative level by the observed expansion, a stalemate was reached regarding which interpretation of the nature of gravity---Einstein's or Eddington's---was supported by the cosmological evidence---as they both used this same observation differently to support their contrasting positions. However, in regard to the essential nature of the expanding Universe itself, \citet{Eddington1933} did have a point, that Einstein's position was explanatorily impotent:
\begin{quote}
There are only two ways of accounting for large receding velocities of the nebulae: (1) they have been produced by an outward directed force as we here propose, or (2) as large or larger velocities have existed from the beginning of the present order of things. Several rival explanations of the recession of the nebulae, which do not accept it as evidence of repulsive force, have been put forward. These necessarily adopt the second alternative, and postulate that the large velocities existed from the beginning. This might be true; but it can scarcely be called an \textit{explanation} of the large velocities\ldots the theory recently suggested by Einstein and de~Sitter, that in the beginning all the matter created was projected with a radial motion so as to disperse even faster than the present rate of dispersal of galaxies, \*(They do not state this in words, but it is in the meaning of their mathematical formulae) leaves me cold. One cannot deny the possibility, but it is difficult to see what mental satisfaction such a theory is supposed to afford.
\end{quote}

Noting Einstein's lack of interest in the cosmological problem after 1917---evident from the facts, that he wrote almost nothing on it afterwards, and that his opinion hardly changed from 1923 onwards---it's not so surprising that he was unconcerned with \textit{why} the Universe should expand. It's clear that the principal reason behind the choices Einstein made in regard to the cosmological problem was that he wanted to reject the $\lambda$-term. And the reason why he was so determined to reject it seems to be that his only concern was to account for a finite density of matter in the Universe. He found that this could be done, with or without the $\lambda$-term, if the cosmic distance scale was time-variable. Therefore, given Hubble's empirical confirmation that cosmic distances \textit{are} time-variable, he could reject the term, which he thought detracted from the logical simplicity of the theory.\footnote{cf. \citet[p.~684-685]{Einstein1949}: ``The introduction of such a constant implies considerable renunciation of the logical simplicity of theory, a renunciation which appeared to me unavoidable only so long as one had no reason to doubt the essentially static nature of space. After Hubble's discovery of the ``expansion'' of the stellar system, and since Friedmann's discovery that the unsupplemented equations involve the possibility of the existence of an average (positive) density of matter in an expanding universe, the introduction of such a constant appears to me, from the theoretical standpoint, at present unjustified.''}

This line of reasoning that Einstein employed in rejecting the cosmological constant is logical enough, yet it seems both na\"{i}ve and short sighted. For one thing, the choice of which formulation of the field equations is ``logically simplest'' is not so easy to determine. On the one hand, in Tolman's opinion, Einstein's choice amounted to an arbitrary and empirically unjustified choice; and in Eddington's, there had been no global consistency in the original theory---$G_{\mu\nu}$ just had to be zero everywhere in empty space, regardless of the form of $g_{\mu\nu}$---so the amended theory was far more simple and elegant.\footnote{For a similar view, cf. \citeapos{Dyson1972} ``Missed Opportunities''.} But on the other hand, in de~Sitter's opinion, the extra term in the equations seemed ``somewhat artificial, and detracts from the simplicity and elegance of the original theory of 1915, one of whose great charms was that it embraced so much without introducing any new empirical constant'' \citep{deSitter1917b}. Einstein certainly sided with de~Sitter in this debate; but it was, in any case, a \textit{debate} over the \textit{interpretation} of a physical theory---a philosophical point by nature---and choosing a side with such absolute certainty, in spite of the fact that Einstein wasn't omniscient, and in fact even lacking any observational evidence to back up his decision, \textit{was} na\"{i}ve. 

If we retrace the above line of reasoning of Einstein's, and don't allow him a hands-down win in the debate over what amounts to a ``logically simpler'' formulation of the field equations of general relativity, the argument already falls apart: he gets the finite density of matter he wants with or without the $\lambda$-term, as long as the Universe is expanding, which Hubble had shown.

He should have been happy. What he seems to have cared the most about was to account for a finite density of matter. But for that very reason, he was already very \textit{unhappy} with the $\lambda$-term, since, although it had allowed him to achieve this aim with his static model---side-stepping the boundary condition problem in the process---de~Sitter immediately used it derive \textit{an empty model that actually satisfies those boundary conditions at infinity}. Immediately after he added the term, it had backfired in the worst possible way, and he wanted rid of it. He cited logical simplicity and the fact that a finite density of matter could be achieved without it, as his arguments against the $\lambda$-term; but surely the fact that an \textit{otherwise satisfactory} vacuous \textit{world} could be described \textit{only if the term existed}, must have influenced his opinion as well. 

Einstein was therefore much happier to employ cosmic expansion as the reason that there can be a finite matter density in the Universe. But then, having found that the Universe is expanding, it was \textit{oddly short sighted} of him to show no concern for \textit{why} the Universe \textit{should} expand. For whatever reason, when it came to the reason for cosmic expansion---which was supposed to be the reason why there can be a finite density of matter!---Einstein seems to have simply given up on his usual philosophical bent. 

Eddington challenged him on this with \textit{The Expanding Universe} in 1933, yet Einstein said nothing more on the cosmological problem until 1945, the year after Eddington died. The timing could have been purely coincidental, and likely had something to do with the Nazi occupation of Germany and the war---but it is unfortunate in any event that he never did take up this debate with Eddington.

Despite efforts like Eddington's to \textit{explain} why our Universe is expanding, interest in the problem slowly waned. Even today, with millions of dollars being put towards ``developing the philosophy of cosmology,'' the grant holders have had no interest in addressing the foundations of relativistic cosmology; cf. \citet{Benetreau2013}. They've concentrated instead on the many concerns that have been realised under the assumption that the basic structure of standard relativistic cosmology is correct,---such as the impossibly rare low entropy state of the big bang \citep{Penrose2005}; or the problems associated with ad hoc proposals such as the inflation scenario, which was proposed in order to `dress up' the standard picture as something it otherwise shouldn't be, in order to ``save the appearances''.

In fact, the irony in addressing only the secondary issues of cosmology, is that they all implicitly assume the same absolute cosmic time that cosmologists---from Einstein and Weyl, to Robertson and everyone else who followed---have always assumed; e.g., which \citet{Einstein1945} implicitly assumed when he wrote that ``The demand for \textit{spatial} [sic] isotropy alone leads to Friedman's form. It is therefore undoubtedly the general form, which fits the cosmologic problem.'' It's only through subsequent recognition that it would be extremely unlikely for space---which is supposed to exist as true cosmic time passes, absolutely---to be inherently isotropic and homogeneous, that the latest problems of interest to philosophers were realised in the first place. In contrast, there's common consensus in the philosophy community that, while relativity isn't strictly incompatible with an absolute time, the assumption \textit{at the very least} amounts to unobservable structure that stands opposed to the theory's fundamental symmetry; see, e.g., \citet{Callender2000}, \citet[ch.~17]{Dainton2001}, and \citet{Maudlin2012}. And the irony of this consensus, is that standard cosmology's assumption of absolute time \textit{has been} empirically confirmed, with increasing precision over the past eighty years. In \S\,\ref{sec:6}, we'll discuss this in more detail---but for now our main interest will remain with the problem of the cosmological constant, which was indeed Einstein's main concern.

Lema\^{i}tre's devotion to the cosmological constant, which he once called ``a second constant demanded by the logical structure of the theory'' \citep{Lemaitre1949b}, has been noted. However, around the time that Lema\^{i}tre was having his conversation with Chandrasekhar, \citet{Pauli1958} published an English version of his earlier relativity textbook, which had a supplemental note on the cosmological problem. After a brief outline of the contributions by Friedmann, Lema\^{i}tre, and Hubble in the 1920s, he wrote, 
\begin{quote}
Einstein was soon aware of these new possibilities and \textit{completely rejected the cosmological term} as superfluous and no longer justified. I fully accept this new standpoint of Einstein's. 
\end{quote}
The steady state theory, which embraced the same description of an intrinsic cosmical repulsion as Eddington's model, may be considered a last-ditch effort that failed when the cosmic microwave background radiation (CMBR) was discovered \citep{Penzias1965,Dicke1965}.---And later yet, in \citeapos{Chandrasekhar1983} centenary ``tribute'' to Eddington (which is also where he recalled the conversation with Lema\^{i}tre), he commented that ``no serious student of relativity is likely to subscribe to Eddington's view that `to set $\Lambda=0$ is to knock the bottom out of space'.''

Probably the most significant empirical result that contributed to this `turning of the tides' was \citeapos{Sandage1958} corrected measurement of Hubble's constant. For, as it turned out, a reasonable order-of-magnitude estimate of the present matter density, taken together with the original measurement of Hubble's constant ($H_0\approx500~\mathrm{km~s}^{-1}~\mathrm{Mpc}^{-1}$, which is roughly an order-of-magnitude too large), gave an estimated age of little more than $10^9$ years \textit{for any Friedmann universe with} $\lambda=0$. It's important that this problem, deduced from the initially erroneous value of $H_0$, was a problem \textit{only if one insisted on setting} $\lambda=0$: with positive $\lambda$, the relation between the present mass density and Hubble's constant does not at once give the age of the universe---e.g. which can decelerate initially, even pausing for an arbitrary length of time near its Einstein radius, depending on the ratio between mass density and $\lambda$, and eventually accelerate towards an asymptotic de~Sitter universe. \citet{Lemaitre1933}, e.g., had long since rejected the ``solutions in which the expansion speed has always been faster than it is now'', and particularly Einstein's ``cycloidal model'', based on this observation. 

Einstein was well aware of the ``time-scale difficulty'' with his favoured $\lambda=0$ models; and elaborating on his position in detail will shed more light on his basic convictions. \citet{Einstein1932} had indeed found that ``it is possible to represent the facts without assuming a curvature of three-dimensional space''---which meant that, given the measured value of Hubble's constant, $H_0\approx500~\mathrm{km~s}^{-1}~\mathrm{Mpc}^{-1}$, they could derive an estimate that agreed, to the correct order of magnitude, with the observed mass density of the Universe, based on a model that assumed no spatial curvature and neglected the $\lambda$-term from the outset. Although they made no mention of the age of such a universe in this paper, it's a simple matter to show that it's given as ${2 \over 3}H_0^{-1}$. Therefore, since they had shown that the Universe is roughly flat---although they noted that the ``curvature is\ldots essentially determinable, and an increase in the precision of the data derived from observations will enable us in the future to fix its sign and to determine its value''---this formula for the age of the universe in terms of $H_0$ remains roughly applicable for realistic curvature values different from zero. Noting that $1~Mpc\approx3\times10^19~\mathrm{km}$, the age of a flat Einstein-de~Sitter universe is roughly $10^9$~years when $H_0=500~\mathrm{km~s}^{-1}~\mathrm{Mpc}^{-1}$.

Einstein discussed this time-scale problem at length at the end of his appendix on the cosmological problem in \textit{The Meaning of Relativity} \citep{Einstein1945}. The order of summarising remarks given in this appendix is in fact significant, as it illustrates Einstein's absolute conviction that the $\lambda$-term should be rejected \textit{even though it would again have provided a solution to the problem he faced}. 

The first point he made, in fact, was that, through Friedmann's solution and Hubble's observation, the $\lambda$-term was to be rejected because it had lost its ``sole original justification---that of leading to a natural solution of the cosmologic problem.'' This was his final mention of the term in the appendix. After that, he noted that the relation between matter density and Hubble's constant that one obtains theoretically after neglecting spatial curvature is confirmed empirically (this was \citeapos{Einstein1932} result). But then he adds: (1) that the age of the universe from the start of expansion is only on the order of $10^9$~years; (2) that this time doesn't concur with the theory of stellar evolution; and (3) that this isn't changed by introduction of spatial curvature or by considering random motions of stars and stellar systems. The rest of his remarks are also concerned with addressing the problem of the brevity of the cosmic lifetime, concluding with a final remark, that 
\begin{quote}
The age of the universe, in the sense used here, must certainly exceed that of the firm crust of the earth as found from the radioactive minerals. Since determination of age by these minerals is reliable in every respect, the cosmologic theory here presented would be disproved if it were found to contradict any such results. In this case I see no reasonable solution.
\end{quote}
As Einstein certainly had to be aware that allowing for positive $\lambda$ would provide a solution, this final sentence must be taken to mean that, in his opinion, invoking $\lambda$ in order to solve the problem would be \textit{unreasonable}.

\citet{Lemaitre1949a} did not see it the same way, as he argued the opposite in his submission to the Schilpp volume containing Einstein's autobiography and his replies to the contributed essays. In response to Lema\^{i}tre's arguments in favour of positive $\lambda$ in the field equations, \citet{Einstein1949} wrote:
\begin{quotation}
\ldots I must admit that these arguments do not appear to me as sufficiently convincing in view of the present state of our knowledge\ldots

The situation becomes complicated by the fact that the entire duration of the expansion of space to the present, based on the equations in their simplest form[ i.e., with $\lambda=0$], turns out smaller than appears credible in view of the reliably known age of terrestrial minerals. But the introduction of the ``cosmological constant'' offers absolutely no natural escape from the difficulty. This latter difficulty is given by way of the numerical value of Hubble's expansion-constant and the age-measurement of minerals, completely independent of any cosmological theory\ldots
\end{quotation}
He'd rather have abandoned relativistic cosmology than invoke the cosmological constant. Einstein clearly saw the introduction of the $\lambda$-term for this purpose as being no more justified than it had been originally. And indeed, in a Lema\^{i}tre-type universe, the $\lambda$-term cannot be justified on the basis that it would provide a natural physical mechanism for expansion, as it would in an Eddington-type universe or in the steady state theory (this is discussed in \S\,\ref{sec:5}, below). Instead, it would provide the same sort of Band-Aid as it had initially, to cover up an exposed wound in the intersection between theory and observation---which is something that Einstein had rightly come to disdain. 

Einstein's conviction that the hypothetical $\lambda$-term should not be resurrected in the same ad hoc manner as it had been proposed initially, in order to reconcile a discrepancy between observation and theoretical expectation, was finally vindicated when \citet{Sandage1958} showed that Hubble's original measurement of the expansion rate was too large by a factor of between 5 and 10. Indeed, as Sandage noted at the end of his paper, ``The major conclusion is that there is no reason to discard exploding world models on the evidence of inadequate time scale alone, because the possible values of $H$ are within the necessary range.'' Unfortunately, the revision of Hubble's constant came three years too late for Einstein to have known.

In fact, a decade ahead of Sandage, while Einstein was still alive, the steady state theory had been proposed, which \textit{also} made prominent use of the cosmological constant. The continued use of the term in Einstein's lifetime was likely a source of irritation for him. In the somewhat oddly unfactual account of early cosmology given in George \citeapos{Gamow1970} autobiography, there is a very famous statement in which Gamow notes that Friedmann's work 
\begin{quote}
opened an entire new world of time-dependent universes: expanding, collapsing, and pulsating ones. Thus, Einstein's original gravity equation was correct, and changing it was a mistake. Much later, when I was discussing cosmological problems with Einstein, he remarked that the introduction of the cosmological term was the biggest blunder he ever made in his life. But this ``blunder,'' rejected by Einstein, is still sometimes used by cosmologists even today, and the cosmological constant denoted by the Greek letter $\Lambda$ rears its ugly head again and again and again.
\end{quote}
This paragraph---which is sandwiched between a description of Einstein's spherical universe as ``stable'', which it's not, and a statement that Hubble ``proved that the so-called spiral nebulae are actually giant stellar galaxies floating in space far beyond the limits of the Milky Way, and the previously observed red shift of the lines in their spectra is to be interpreted as a result of their mutual recession'' \textit{all in the same year} that \citet{Friedmann1922} published his first paper, which is also wrong---illustrates as well the change of heart that was taking place within the physics community during the latter half of the twentieth century, when the Einstein-de~Sitter expansion paradigm finally began to set.  

As noted above, the next significant strike against the cosmological constant, after Einstein's rejection of the $\lambda$-term based on the discovery of expansion, was \citeapos{Sandage1958} discovery, which showed that the Einstein-de~Sitter model was not actually ruled out by empirical data. The paper by Sandage actually begins by stating, ``In principle, a decision between the simplest cosmological models (exploding cases with $\Lambda=0$, $k=+1,0,-1$, or the steady-state case) is possible from the observed velocity-distance relation.'' And later on he wrote that ``The simplest theory is the general theory of relativity together with Friedman's exploding models with $\Lambda=0$.'' Therefore, the tides seem finally to have begun to turn,---and not just away from the idea that cosmic expansion should be essentially driven by the cosmical repulsion due to $\lambda$, as Eddington had advocated, but from use of the $\lambda$-term in general.

From the perspective of cosmology, the final blow came when the steady state theory was ruled out by \citeapos{Penzias1965} discovery of the big bang's relic radiation, which the theory couldn't naturally account for---and, more importantly, which was a prediction of the big bang theory \citep{Gamow1946,Alpher1948a,Gamow1948,Alpher1948b,Alpher1950}.

As the above statements by Sandage indicate, Einstein's rejection of the $\lambda$-term had already contributed to a significant shift away from the original concept of cosmic expansion, through which the idea was initially accepted; and the subsequent empirical measurements that showed again and again that the motivation for its re-introduction was indeed false and unjustified, led to a general renunciation of the term. 

At the end of the century, observational cosmology was commonly practised using the same two-parameter model discussed by \citet{Sandage1958}, involving Hubble's constant, which describes the present velocity of expansion, and a deceleration parameter that would describe just how rapidly the ``Einstein-de~Sitter universe'' was slowing in its expansion. The idea that the Universe has been slowing down since it began---which \citet{Eddington1933} once decried for its explanatory impotence and \citet{Lemaitre1933} had shown (incorrectly) to be inconsistent with the empirical data---was hardly in question. 

In any case, when the cosmological constant was considered, the reason had nothing to do with an \textit{expectation} that $\lambda$ should be positive in an expanding universe; see, e.g., \citet{Carroll1992} and \citet{Goobar1995}. It was therefore quite unexpected when in 1998 the first concrete evidence was found to indicate that the deceleration parameter is negative (the cosmic expansion is accelerating), and that the variable rate of expansion is \textit{consistent with the presence of a cosmological constant} \citep{Riess1998,Perlmutter1999}. Thus, a few years later \citet{Ellis2003} referred to the discovery from an historical perspective as ``the triumph of an unwanted guest''. And \citet{Penrose2005} stated, ``For my own part, in common with most relativity theorists, although normally allowing for the possibility of a non-zero $\Lambda$ in the equations, I had myself been rather reluctant to accept that Nature would be likely to make use of this term. However\ldots, much recent cosmological evidence does seem to be pointing in this direction.'' 

In fact, a positive-valued cosmological constant was so unexpected---so \textit{unwanted}---from a theoretical standpoint, apart from a possibility that it could be connected to an unexplained vacuum energy calculation from quantum physics that's off by a hundred orders of magnitude, and might be fine-tuned to the observed value \citep{Weinberg1989,Carroll1992}, that the cosmological constant, having finally been measured to have a positive value, was commonly reconceived as ``dark energy''; e.g., \citet{Peebles2003} wrote:
\begin{quote}
We shall use the term \textit{dark energy} for $\Lambda$ or a component that acts like it. Detection of dark energy would be a new clue to an old puzzle: the gravitational effect of the zero-point energies of particles and fields. The total with other energies, that are close to homogeneous and nearly independent of time, acts as dark energy. What is puzzling is that the value of the dark-energy density has to be tiny compared to what is suggested by dimensional analysis; the startling new evidence is that it may be different from the only other natural value, zero.
\end{quote}
Thus, the observed cosmological constant was immediately shanghaied, without a thought to the possibility that it could be a fundamental constant, \textit{essential} to the existence of such an expanding universe as ours---i.e. without a thought that the concept of expansion that held sway prior to observation of this ``unwanted guest'', through which the cosmological-term had been commonly dismissed, \textit{could be fundamentally flawed}. 

Over the past fifteen years, little progress has been made towards a better understanding of the nature of the supposed dark energy; and despite proposed modifications, the best model consistently appears to have a pure cosmological constant \citep{Riess2004,Riess2007,Davis2007,Hicken2009,Riess2011,Suzuki2012,Hinshaw2013,Ade2013}. The dark energy hypothesis does not subscribe to Eddington's idea that without a cosmological constant there's really no reason why the Universe should expand in the first place, but remains in the spirit of Einstein and de~Sitter \citep{Eddington1933}:
\begin{quote}
I find a difficulty in discussing the proposal of Einstein and de~Sitter, and some of de~Sitter's separate proposals, because I do not see what are ``the rules of the game''. These proposals are left as mathematical formulations, all doubtless compatible with what we observe; but there seems nothing to prevent such formulations being infinitely multiplied. De~Sitter has several times emphasised the possibility that the cosmical constant $\lambda$ might be negative. That gives cosmical attraction instead of repulsion. Clearly the recession of the nebulae is not evidence in favour of cosmical attraction. The most that can be said is that it is not necessarily fatal evidence against it\ldots

I might discuss these suggestions more fully if they were likely to be the last. But it would seem that, unless we keep to a defined purpose, an unlimited field of speculation is open; and by the time these remarks are read, some other hypothesis may be in vogue.
\end{quote}

The time is ripe for us to begin questioning the foundations of our thinking about the one and only Universe we live in, and about the cause of its expansion in particular; for, as \citet{Ijjas2013} recently remarked, ``In testing the validity of any scientific paradigm, the key criterion is whether measurements agree with what is expected given the paradigm''. Cosmic expansion is allowed, but isn't something that should ever be expected from an Einstein-de~Sitter point of view; and the observed cosmological constant really is an ``unwanted guest'', as our discussion of Einstein's cosmological considerations has shown. In contrast, the unexpected and unwanted measurement that was made \textit{in our Universe} is in fact precisely what would have been expected if the superseded idea of expansion due to cosmical repulsion had prevailed instead. The foundational considerations in cosmology should be revisited.

\section{What possible way forward?}
\label{sec:5}
What lessons are to be learned from the history of ideas in modern relativistic cosmology? Particularly as regards the cosmological constant, a number of points can be drawn from the ups, downs, and side steps it's taken this past century. By considering these, we'll look for a possible way forward.

Perhaps the main point to be learned, is that it is simply bad practice to invent new hypotheses of an ad hoc nature, just to ``save the appearances'' within a given theoretical framework. The classic example that stands against this is the addition of epicycles to account for retrograde motion in the geocentric model of our solar system, and the argument that this is all right to do because all that matters is to accurately \textit{describe} observations, saving the appearances by whatever means available; that the true nature of Physical Reality is not for us to determine. Newton's law of gravitation, which came as a direct consequence of the efforts made by people like Copernicus, Bruno, Galilei, and Kepler, to improve our basic understanding of Nature, is an example of the type of advance that \textit{can} be achieved by questioning the foundations of a failing system of thought, rather than slapping Band-Aids all over it in perpetuity. In contrast, the hardcore empiricist stance,---that accurately describing the facts is the \textit{only} thing that matters, so long as the appearances are ``saved''---is \textit{against} progress.

However, Einstein initially proposed the $\lambda$-term for this very reason, i.e. in order to reconcile two basic assumptions that he considered necessary---viz.\ that the world-structure should be static, as the evidence seemed to indicate, and that this structure should correspond to a finite density of matter, in accordance with certain philosophical predilections that had guided the formulation of his general theory of relativity. When it became obvious that the world-structure is not static, Einstein rejected the cosmological-term, not because it was incompatible with time-variable world structure, but because the ad hoc supplementation of the basic general relativistic framework had proven both unnecessary from the original standpoint \textit{and} not strictly in line with the assumption about the overall world-structure being determined by a finite matter density, as \citeapos{deSitter1917a} result had shown.

But then \citet{Lemaitre1933} soon realised that actually setting  $\lambda=0$ as Einstein did, had the negative effect that, when taken in conjunction with the observed matter density and expansion rate of the universe, it led to an estimate of the age of the universe that was less than the known age of the Earth. However, while Lema\^{i}tre argued that the problem could be avoided by invoking $\lambda$, this time Einstein would not budge, as he effectively stated that he'd rather just abandon relativistic cosmology altogether than make the ad hoc assumption. And again, better empirical evidence did show that there actually was no problem, as it turned out that the measured expansion rate was too high by an order of magnitude, and the age-estimate was consequently too low by an order of magnitude; therefore, the Universe could be older than the Earth without the need for $\lambda$.

Still, this was not the last proposal that non-zero $\lambda$ might be indirectly evidenced by the data. In 1967, the suggestion was made yet again, this time based on an apparent preponderance of quasars at $z=1.95$ \citep{Petrosian1967,Shklovsky1967,Kardashev1967}. The idea was that this observation could be explained if we lived in a Lema\^{i}tre universe that decelerated nearly to a halt at its ``Einstein radius,'' paused there for a very long time so that light previously emitted by distant quasars would develop a prominent absorption spectrum as it passed through the intergalactic medium of quasi-static space, so that, when the cosmological constant finally tipped the scale and exponential expansion ensued, the spectra of all these quasars would appear redshifted by the same amount (presumably $z=1.95$). As it turned out, there is no ``$z=1.95$ phenomenon'': the number of known quasars has increased by four orders of magnitude since 1967, and with that a picture has emerged that describes quasar evolution as continuous, as we've now observed it to be out to $z>6$. Therefore, this hypothesis was ruled out, as with the two earlier proposals of Einstein and Lema\^{i}tre, based on improved empirical knowledge. 

The lesson has come down to us, time and time again, that when we've realised some discrepancy between our basic theoretical framework and the data of observation, it is very bad practice to attempt to reconcile the two by making the slightest alteration to the theory's foundations, if that would serve no explanatory purpose; for when these slight changes have been made merely in order to save the appearances, they simply haven't worked out: we end up with descriptions that \textit{do} evidently work, but there is no reason why they \textit{should} work. The reason why this practice is doomed to fail should be obvious: the basic theoretical framework constitutes all of our basic assumptions about physical reality---what we think should be, based on all that we've observed;---therefore, when we realise a discrepancy between what we think should basically be happening and what evidently is happening, the indication really must be either that the evidence is wrong (e.g., as in the above ``strike-out'' with the cosmological constant) or that something's amiss in what's supposed to be happening (e.g., as with geocentrism). To use \citeapos{Forscher1963} metaphor: when the great edifice turns out all misshapen and crumbly, it must be due either to a poor foundation, or else some bad bricks got added along the way.

Historically, the geocentrists believed that everything went around the Earth; but because the planets periodically pause in their motion with respect to the fixed stars, move backwards for a while, and then resume their course, the geocentric framework had to be supplemented with epicycles. Einstein believed the finite density of matter should determine cosmic structure; but because the universe appeared to be approximately static, he supplemented the relativistic framework with the $\lambda$-term. In both cases, the appearances were saved well enough for a time, and in both cases no reason was given for altering the fundamental theoretical framework in the way that was done, other than a noted discrepancy with observation; and these two cases provide perfect examples of the two things that go wrong when the foundations are altered with no explanatory aim.

In the case of geocentrism, the reason for the discrepancy was a fundamental flaw in the theoretical framework---i.e., our solar system is heliocentric, rather than geocentric---and the Keplerian system provides a very natural explanation of the retrograde motion. In Einstein's case, the evidence later showed that the universe isn't static---i.e. the flaw was empirical rather than fundamental---so the theoretical framework and the data could be reconciled without being supplemented. In both cases, the theoretical framework was compromised because it evidently conflicted with physical reality; physical reality was supposed to be a certain way, but it evidently couldn't be just like that, so a compromise was made whereby it could be as near as possible to what it was supposed to be while also saving the appearances. However, a theory should never be considered complete until it is exactly what it's supposed to be, without having to be supplemented by any epicycles or constants that need to be constructed ad hoc, just to save the appearances.

Despite providing a perfectly accurate description of the data, the standard model of cosmology simply isn't what it's supposed to be; it's all description with no explanation. One of the most well known problems in cosmology is the accelerated expansion of the Universe that's been evident now for fifteen years \citep{Riess1998,Perlmutter1999}. According to the current paradigm, there is simply no good reason why this should be so---yet through this observation we do finally know, with as much confidence as we've ever had about the retrograde motion of the planets, that despite $\lambda$'s previous strike-out, this acceleration is due to something that acts very much like a cosmological constant. 

Even so, the prevailing opinion these past fifteen years has been an Einsteinian one, which assumes above all that the structure and evolution of the Universe is determined by its material content, and therefore takes the observation as an indication that our Universe possesses a ``dark energy'' component that's driving this acceleration. One of the main tasks in observational cosmology has therefore been to constrain the nature of this dark energy. And the outcome of this effort, with better and more data coming in steadily, has consistently been that the dark energy acts \textit{just} like a cosmological constant \citep{Riess2004,Riess2007,Davis2007,Hicken2009,Riess2011,Suzuki2012,Hinshaw2013,Ade2013}.

From an Einsteinian perspective, this is the worst possible outcome, as it turns out that something like Einstein's epicycle really is needed to account for the evidence. If it had turned out that the dark energy was anything but a pure cosmological constant, there might have been some reason to keep pushing the old envelope,---i.e. assuming that the world structure should be a product of its finite matter densities, and that our Universe should really be just like Einstein envisioned it when he described his statical world in 1917, supplementing the field equations of general relativity with the $\lambda$-term because that was the only possible way to uphold his vision.

But the Universe really is very, very much different from what Einstein envisioned. For Einstein always had it in mind, that the average finite matter density of the Universe should determine its structure, and this was most naturally conceived in a static universe model. The hitch was that, just as Ptolemy had required epicycles to uphold his world-view, Einstein had required the $\lambda$-term to uphold his. The cosmological constant was a blemish, and Einstein was glad to reject it in favour of cosmic expansion, which was an empirically motivated means to serve \textit{the very same purpose}. As \cite{Einstein1932} put it, 
\begin{quotation}
Historically the term containing the ``cosmological constant'' $\lambda$ was introduced into the field equations in order to enable us to account theoretically for the existence of a finite mean density in a static universe. It now appears that in the dynamical case this end can be reached without the introduction of $\lambda$.
\end{quotation}
The original ``epicycle'' was simply replaced with one that agreed with the evidence. Yet the troubling thing about ``epicycles'' is that, regardless of empirical motivation, there's no basic reason why they \textit{should} be; indeed, there is no reason why a universe like the one Einstein envisioned should ever expand. And now this great explanatory weakness of the theory has indeed been amplified, since the evidence has come to support cosmic expansion in a form that agrees with the existence of a cosmological constant.

Therefore, while Einstein does deserve some credit for his eventual refusal to use the $\lambda$-term as an ad hoc device, as noted previously his ultimate neglect to search for any explanation of expansion is worthy of reproach. For in fact, in rejecting the $\lambda$-term, Einstein also rejected the only known interpretation of expansion that would explain why the universe should expand, as his interest in cosmic expansion only extended so far as \textit{that} could be used to support his principal bias, that the large scale structure of the universe should be determined by its finite matter density. Einstein's attitude in this matter was uncharacteristically irrational, as he was content to leave the fact of the Universe's expansion entirely unexplained.

And this great omission has forever haunted Einstein's theory. For indeed, modern cosmology continues to be based on the principle that the large scale structure and evolution of the Universe should be determined by its matter density, and the problems with this picture keep piling up. Most recently, we've come to find that 95\% of that matter can't be the kind that Einstein wanted the Universe filled with in the first place---and that the great majority of that should in fact be something that acts just like a cosmological constant. As the evidence from cosmology has amassed, more and more has been heaped onto an Einsteinian expansion paradigm without ever going back to question whether the assumptions of that paradigm remain valid in light of today's evidence. And the universe of today's standard cosmological model, with its inflaton, dark matter, and dark energy fields, indeed looks nothing like Einstein's vision of what the ultimate large scale application of general relativity should be---in particular, the dark energy field, which hasn't shown a sign of being anything but a positive cosmological constant, stands directly opposed to the Einsteinian paradigm, as foundational evidence of the alternate interpretation of expansion that Einstein abhorred.

As discussed above, when the time comes that the theoretical framework needs to be supplemented with ``epicycles'' just to ``save the appearances'', there is a clear indication that either the appearances themselves are wrong, or the fundamental theoretical framework is flawed. And in this case, it seems that both of Einstein's epicycles---the cosmological constant and the expansion of the universe---are as clearly motivated by the evidence as Ptolemy's epicycles were---which we can therefore reasonably take as an indication that the theoretical framework is flawed; i.e. that relativistic cosmology's standard model is fundamentally flawed. 

As noted at the start of this paper, the description of cosmic expansion that's given by the flat $\lambda$CDM model agrees extraordinarily well with the data, and the problem is that we don't know \textit{why} the Universe should be the way it appears---i.e. why it should be flat, with its expansion rate mainly influenced by $\lambda$ and \textit{cold dark matter}; or why it is apparently maximally non-local, with the cosmological horizon, defined by light that has travelled 13.8 billion years through space only to reach us now, indicating that the same thing was going on 13.8 billion years ago in every direction that we look, in regions that \textit{should} have been---and should still be---causally disconnected. In fact, there is no reason why such a universe should ever have come to expand at all. Therefore, because the flat $\lambda$CDM model works perfectly well despite the fact that we've got no idea why it should work, it shouldn't be at all surprising if, instead of having to replace the cosmological constant, or cosmic expansion for that matter, with something else---e.g., as Ptolemy's epicycles were replaced by a heliocentric system of orbits---what's needed is to find a different theory that leads to the exact same description, and through which the ``epicycles'' of the standard cosmological model should be interpreted differently.

And indeed, in light of all the evidence that's accumulated in the century since Einstein first proposed the $\lambda$-term, it seems that \textit{Eddington's} interpretation of $\lambda$, rather than Einstein's, should be particularly promising. For one thing, all the cosmological evidence does now indicate that there is a cosmological constant. Indeed, even from a purely kinematical perspective, it's since been noted that $\lambda$ carries Minkowski's (1908) argument ``to its logical conclusion'', whereby the most general possible kinematical group is described, which degenerates to the Poincare group of special relativity in the flat-space limit \citep{Bacry1968,Dyson1972}. And then, more to the point, as Eddington noted, the natural cosmical repulsion described by the $\lambda$-term really \textit{could} explain why our Universe should expand. Indeed, as \citet{Dyson1972} commented, ``Suppose that somebody had been bold enough in 1908 to take [the argument from kinematics] seriously. He would have correctly predicted the expansion of the universe twenty years before it was discovered observationally by Hubble.'' And so it was, as we've already noted, that citet{Eddington1933} viewed the expansion of the universe as ``Einstein's almost inadvertent prediction.''

But there is a snag in this hope that the cosmical repulsion due to $\lambda$ could explain expansion, and simply reconceiving things won't do---for if $\lambda$ were the basic mechanism of expansion, it would have to drive expansion for all time; and indeed, Eddington always avoided the big bang models, referring to them as ``too unaesthetically abrupt'', for a good reason:---all of the Friedmann-Lema\^{i}tre-Robertson(-Heckmann)-Walker (FLRW) models that expand from a point with zero proper volume at a finite time in the past \textit{have to be} decelerating at the outset; while a cosmological constant drives the cosmic length scale $R$ (commonly known as the \textit{scale-factor}) to accelerate at a rate proportional to $R$, regular non-relativistic matter causes these universes to \textit{decelerate} at a rate proportional to $R^{-2}$ and radiation causes deceleration as $R^{-3}$; therefore, as $R$ approaches zero, the contribution from cosmic repulsion becomes \textit{insignificant}, while the contributions from matter and radiation \textit{blow up} exponentially. Even the flat $\lambda$CDM model that fits the cosmological data so well \citep{Hinshaw2013,Ade2013}, which comes to accelerate as matter and radiation densities become small enough,---must be decelerating at the beginning, with negligible influence from the cosmological constant. The cosmical repulsion cannot be the \textit{cause} of expansion \textit{in these universes}.

In fact, there is no physical mechanism to explain why \textit{these} big bang universes would ever expand, since the cause of expansion can \textit{only} be attributed to a singularity where the whole physical theory blows up \textit{and in the vicinity of which we do know that physical effects actually} oppose \textit{expansion}. 

It's important to properly appreciate this particular aspect of our standard model. Consider, e.g., Coulomb's law: 
\begin{equation}
\mathbf{F}=k\frac{q_1q_2}{r_{21}^2}\hat{\mathbf{r}}_{21}.
\end{equation}
Here, the electrostatic force between two charges $q_1$ and $q_2$ is proportional to the inverse-square of the distance between them. Depending on whether the charges have the same or opposite signs, the force will be repulsive or attractive, respectively. In either case, the law blows up when the distance goes to zero, so the physics simply doesn't work when two charges occupy the exact same position---but a consistent interpretation is that the repulsion or the attraction becomes infinite as the distance between the charges becomes zero.

In the FLRW models, the \textit{decelerative} gravitational effect, through which the universes should tend to collapse rather than expand, also blows up to infinity at the big bang. Near the big bang, that decelerative force is enormous, and precisely at the big bang, the physics completely breaks down. But while the physics breaks down there, the most reasonable inductive inference we can make from the continuous limit in its near vicinity is that the big bang singularity should be a point of infinite attraction. Not only does the model provide no indication why these universes \textit{should} ever expand; it actually strongly indicates that they \textit{shouldn't}.

If we take the best model for our expanding Universe at face value as a description of Physical Reality that does more than just ``save the appearances'', but Really captures what's actually going on, the indication is that our Universe \textit{shouldn't} be expanding; it never \textit{should} have come into existence! This is the explanatory impotence of the theory that \citet{Eddington1933} couldn't stand.

There is no good reason to say that such a singularity should be the \textit{cause} of cosmic expansion---and indeed, of all existence!---except to say that the model fits the data and this is the only way we know to interpret it. In that case, physical cause gets completely swept under a perfectly non-physical rug, and physics itself is reduced to a purely empirical science \textit{based on a non-physical axiom}. 

Hoyle's term for the ``big bang'' hypothesis is perfectly appropriate; a horrible dilemma is reached:---either we use anthropic reasoning against inductive reason, to suggest that a mystical power acted against physics at the first instant to produce the universe our model describes, taking that description at face value; or we admit that while our model is empirically accurate, there may be a fundamental flaw either in the physics or in our interpretation of it. The way forward for the physicist or philosopher must surely be the second---to investigate new physics or new ways of interpreting the physics, that wouldn't suffer this evident flaw;---but while that path is surely more difficult than becoming a member of the Church of the Big Bang Singularity, which teaches belief in non-physical axioms, it should be of some consolation to the hard-pressed philosopher-scientist to know that there \textit{is} an empirically evident and theoretically motivated physical mechanism that would drive expansion---viz. the cosmical repulsion due to $\lambda$---which can serve a heuristic purpose in that pursuit.

As \citet{Ijjas2013} noted, disagreement between measurement and expectation within a given scientific paradigm is the key criterion on which the paradigm is based. A scientific paradigm is more than just a mathematical model, but consists of all the suppositions that enter into the model and inferences that are made regarding measurement constraints---i.e., what the mathematical model is thought to mean.  We have, at present, a mathematical model---the flat $\lambda$CDM model---that fits the data remarkably well,---in fact, better than proposed alternatives that have thusfar been based on hypothetical modifications to physical theory \citep{Riess2004,Riess2007,Davis2007,Hicken2009,Riess2011,Suzuki2012,Hinshaw2013,Ade2013}. Even so, apart from the CMBR, none of the cosmological observations have been as expected, according to the paradigm: the matter that the Earth, stars, nebulae, galaxies, etc. are composed of, which Einstein fought to describe as a fundamental constituent of the world, is now known to account for only about five-percent of the energy density described by the model; the curvature is either zero, or very nearly that, which is the most unlikely possibility out of the FLRW universes that theoretically could have been, and we've invented inflation in an ad hoc attempt to flatten things out, which seems not to have worked out \citep{Steinhardt2011,Ijjas2013}; inflation also attempts to reconcile our expectation that cosmic structure should have emerged as a local phenomenon limited by the finite speed of light, with the fact that without a mechanism like inflation, observations out to our cosmic event horizon indicate that our Universe is actually maximally non-local; and even cosmic expansion itself, we've seen, is something that should be wholly unexpected within the current paradigm.   

The problems of cosmology, it would seem, have little to do with our ability to construct a mathematical model that \textit{describes} observations. On the contrary, the observationally constrained model fits the data very well, and is in many ways the simplest possibility that could have been, despite the numerous complications that have been proposed. According to this model, the evolution of the scale-factor is a simple trigonometric function parametrised only by $\lambda$, as (see, e.g., \citet{Janzen2012a,Janzen2012b})
\begin{equation}
R(t)\propto\sinh^{2/3}\left(\frac{3}{2}\sqrt{\frac{\lambda}{3}}t\right).
\end{equation}
Instead of complicating the mathematics, it stands to reason that our expectations may be inconsistent with the description given by this simple model because our approach to the problem, and therefore our interpretation of the observational results, has been wrong. 

A possible way forward for cosmology therefore seems to be to re-examine the basic assumptions of the standard model, keeping in mind that the cosmological-term in the Einstein field equations, with a positive cosmological constant, describes cosmical repulsion, and that the best model we've found to describe the evolution of the scale-factor of our Universe is effectively parametrised by non-vanishing $\lambda$ only. 

By re-evaluating those basic assumptions now, in light of the empirical evidence that's accumulated through the past century, it may well be possible to develop a fundamentally different theory of our expanding Universe that agrees as well with observation, perhaps differing only by reinterpretation of the constrained model.

\section{A possible way forward}
\label{sec:6}
The basic ingredients of the standard big bang cosmological model are general relativity theory (GRT) and the Robertson-Walker (RW) line-element
\begin{equation}
ds^2=-dt^2+R(t)^2d\sigma^2,
\end{equation}
wherein $d\sigma^2$ describes isotropic and homogeneous space. To this basic framework, we add certain elements like an inflaton field, which is supposed to flatten things out and causally connect our current observational horizon at a time prior to inflation, as well as appreciable dark matter and dark energy fields, which are the primary influences on scale-factor evolution throughout most of the universe's history. This description works, although there's no good reason why the type of inflation that's allowed by the data should actually have taken place, why there should be any dark matter or dark energy, or why the universe, as described, should ever have come to expand at all.

As we'd like to avoid having to add anything at all to our basic framework, we can focus our attention on determining what might be wrong with its GRT+RW basis, as indicated by our need to supplement it with all this extra stuff. While it's generally acknowledged that some modification, at least to our basic understanding of general relativity, if not to the mathematical formalism itself, will be required in order to reconcile the theory with quantum physics, we can note right away that whatever this is, it's not likely to help with our principal concern of wanting to explain why our Universe should expand---why we should theoretically expect it to expand. 

The reason that the widely sought theory of quantum gravity isn't likely to help with our problem, is that it should only be relevant to cosmology in the first moment, and there's no reason to expect that any modification to theory there should explain why a universe filled with, and for billions of years primarily influenced by, stuff that would work to keep it from expanding, should ever come to expand rather than remain tightly bound or contract. Given that the decelerative force increases exponentially towards $R(t)=0$, in all likelihood the quantum-gravitational modification of that point won't describe an enormously repulsive Planck-scale moment.

This significant point was indeed overlooked by \citet{Einstein1945}:
\begin{quotation}
The theoretical doubts [about the assumption of a ``beginning of the world'' (start of the expansion)] are based on the fact that for the time of the beginning of expansion the metric becomes singular and the density, $\rho$, becomes infinity\ldots For large densities of field and of matter, the field equations and even the field variables which enter into them will have no real significance. One may not therefore assume the validity of the equations for very high density of field and of matter, and one may not conclude that the ``beginning of the expansion'' must mean a singularity in the mathematical sense. All we have to realize is that the equations may not be continued over such regions.
\end{quotation}
The problem has little to do with the description of the initial point \textit{as} a singularity: it's that the first instant, and whatever actual physics that describes it, must explain why there \textit{should} subsequently be an expanding universe, which all the well-defined physics actually opposes. Whatever the actual physics of such an initial singularity as the big bang models describe, it's unlikely to explain the existence of an expanding universe.

This leaves the option of scrutinising the basic assumptions of the RW line-element, which can be read off from the fourth page of \citeapos{Robertson1933} unified account of relativistic cosmology. The first two are: i. an absolute division of space and time, with absolute space (i.e. spacelike surfaces of absolute or ``true'' simultaneity) evolving (i.e. expanding, etc.) through the course of absolute, cosmic time, $t$, which is given as the proper time of a congruence of geodesics; and ii. a further requirement that these fundamental world lines are orthogonal to the spaces $t=const.$ Therefore, as \citet{Robertson1933} notes, the synchronous sets of events given by hypersurfaces of constant cosmic time ``reinstate an absolute simultaneity into the actual world''---i.e. that's in agreement with \citeapos{Einstein1905} operational definition of ``simultaneity''---and ``allows us to give a relatively precise formulation of the assumption that our ideal approximation to the actual world is spatially uniform.'' The other two assumptions, then, are that this objectively defined space should be iii. isotropic and iv. homogeneous.

Two of these assumptions (i.; at least its kinematical content) and (iii.) are supported by observation; one (iv.) is the mathematical expression of a principle no less fundamental than the principle of relativity \citep[\S\,I.2]{Einstein1905}; and the other (ii.) comes from an assumption on the ontological meaning of relativity, and particularly the meaning of the relativity of simultaneity \citep[\S\,I.1]{Einstein1905}.

Assumptions iii. and iv. require little remark: the Universe, as we observe it, has been isotropic on the large scale for all time; and, although we can't observe snapshots of constant time due to the finite speed of light, we can reasonably infer that this observation should be the same regardless of location within the Universe. By the observation of large scale isotropy and the \textit{cosmological principle}, the Universe at all times should be isotropic and homogeneous. But this already assumes a cosmic time throughout which this three-dimensional space can be described to exist and evolve; hence, the prior requirements i. and ii.

\citet{Robertson1933} identified i. and ii. with Weyl's postulate, but it must be clear that Einstein himself had a great deal of influence on the introduction of these assumptions into physical theory. For one thing, we've noted that Einstein was the first to take the step, in relativistic cosmology, back towards a description of the universe as absolute three-dimensional space that exists (evolves, endures) in the course of absolute time. This was one of \citeapos{deSitter1917a} immediate criticisms of \citeapos{Einstein1917} model, which we've discussed in \S\,\ref{intro}. Later that year, \citet{deSitter1917b} wrote,
\begin{quote}
It thus appears that the system A [Einstein's] only satisfies the mathematical postulate of relativity if the latter is applied to three-dimensional space only. In other words, if we conceive the three-dimensional space ($\mathrm{x}_1,\mathrm{x}_2,\mathrm{x}_3$) with its world-matter as movable in an absolute space, its movements can never be detected by observations: all motions of material bodies are relative to the space ($\mathrm{x}_1,\mathrm{x}_2,\mathrm{x}_3$) with the world-matter, not the absolute space. The world-matter thus takes the place of the absolute space in Newton's theory, or of the ``inertial system.'' It is nothing else but this inertial system materialised. It should be pointed out that this relativity of inertia is in the system A only realised by making time practically absolute. It is true that the fundamental equations of the theory\ldots remain invariant for all transformations. But only such transformations for which at infinity $t'=t$ can be carried out without altering [the field at infinity]. In the system B [de~Sitter's], on the other hand, there is complete invariance for all transformations involving the four variables.
	
The system B is the four-dimensional analogy of the three-dimensional space of the system A\ldots
	
In both systems A and B it is always possible, at every point of the four-dimensional time-space, to find systems of reference in which the $g_{\mu\nu}$ depend only on one space-variable (the ``radius-vector''), and not on the ``time.'' In the system A the ``time'' of these systems of reference is the same always and everywhere, in B it is not. In B there is no universal time; there is no essential difference between the ``time'' and the other three co-ordinates. None of them has any real physical meaning. In A, on the other hand, the time is essentially different from the space-variables.
\end{quote}
Thus, de~Sitter immediately called Einstein on not being a true relativist, as he'd assumed an absolute time in the development of his cosmological model. \citet[p.~150]{Eddington1920} noted the same, but without the same air of criticism: 
\begin{quotation}
[It may be objected that in Einstein's cosmological model] absolute space and time are restored for phenomena on a cosmical scale. The ghost of a star appears at the spot where the star was a certain number of million years ago; and from the ghost to the present position of the star is a definite distance---the absolute motion of the star in the meantime. The world taken as a whole has one direction in which it is not curved; that direction gives a kind of absolute time distinct from space. Relativity is reduced to a local phenomenon; and although this is quite sufficient for the theory hitherto described, we are inclined to look on the limitation rather grudgingly. But we have already urged that the relativity theory is not concerned to deny the possibility of an absolute time, but to deny that it is concerned in any experimental knowledge yet found; and it need not perturb us if the conception of absolute time turns up in a new form in a theory of phenomena on a cosmical scale, as to which no experimental knowledge is yet available. Just as each limited observer has his own particular separation of space and time, so a being coexistive with the world might well have a special separation of space and time natural to him. It is the time for this being that is here dignified by the title ``absolute''\ldots Some may be inclined to challenge the right of the Einstein theory, at least as interpreted in this book, to be called a relativity theory. Perhaps it has not all the characteristics which have at one time or another been associated with that name\ldots
\end{quotation}

While Einstein never offered further rationalisation for the assumption of an absolute time in relativistic cosmology, beyond what was given in his 1917 paper, it is significant that the later models he preferred---the Friedmann model he favoured in 1931, and the general FLRW models with $\lambda=0$ that he preferred when he learned that the constant curvature of space could be zero or negative---all define the same absolute separation of space and time. Indeed, in the appendix to \textit{The Meaning of Relativity} that he added on the cosmological problem, he included a detailed ``intuitive'' derivation of the description of time-varying rotationally symmetric space \citep{Einstein1945}. By sheer perseverance with these models, Einstein admitted that he was no true Leibnizian ``relativist''. His was simply an empirically Lorentzian world, in which the relativity of simultaneity is really a farce.

But far from chastising him on this account, as de~Sitter did, his foresight should be commended---as it turns out there is strong empirical evidence to support reneging on the overzealously anti-Newtonian meaning of relativity that he originally assumed \citep{Einstein1905}. We've already noted (in \S\,\ref{intro}) that his original justification for doing so, although it turned out to be wrong, can be reconceived from the fact that the proper velocities of all the galaxies we do observe are very small compared to the speed of light. In a true four-dimensional relativistic world---in which time doesn't \textit{really} pass, simultaneity is \textit{truly} relative (i.e. there are no ``privileged'' observers), and ``inertial motion'' is purely random (see, e.g., \citet{Putnam1967})---observable velocities should be uniformly distributed up to the speed of light. The observation that this isn't so---i.e. that the relative velocities of galaxies through space are all small compared to the speed of light---is arguably the weakest bit of empirical evidence that has been accumulated, since the advent of relativistic cosmology, against the dictum, passed down since the time of Newton and Leibniz, that absolute motion can't be detected. But the fact that proper motions of galaxies \textit{are} small compared with \textit{c}---as we infer them to be in our expanding universe---\textit{is} empirical evidence of a cosmic state of rest.

The second bit of cosmological evidence that suggested this ``causal coherence'' of fundamental world lines was the redshift-distance relation; and Weyl's acknowledgement of the fact \citep{Weyl1923a,Weyl1930} was why \citet{Robertson1933} wrote that it had been ``emphasized above all by Weyl''. The reason for this is straightforward: if the redshift-distance relation is to be attributed to a divergence of world lines in an expanding universe, i.e. describing a three-dimensional association of bodies that gradually disperse due to the expansion of space in which they all remain roughly at rest throughout the course of time, then the assumption that it is \textit{space} that \textit{exists}, enduring as it expands throughout the course of time, needs to be made a priori. For only then is there any meaning in the statement that ``the universe is expanding'', or that it is 13.8 billion years old (despite the fact that only minutes of proper time may have elapsed for some particles that have been travelling through it all that time, with absolute motion near the speed of light through absolute space). More than anything else, it was the observed velocity-distance relation that gave convincing evidence of an objective separation between space and time; and it was our need to describe it that led the relativists of the 1920s and 1930s to a renunciation of the prior relativistic dictum that there can be no absolute space, time, and motion.

But today, it's the CMBR that provides us with overwhelming evidence that there is an absolute cosmic rest frame. For the CMBR was predicted, first and foremost, upon the basis that shortly after the big bang, at a particular point in time when three-dimensional space was relatively much more dense, all the matter would have been an opaque plasma that eventually cooled enough for atoms (viz.\ Hydrogen, as heavier nuclei wouldn't have had any chance to form before then) to form, and then decouple from radiation, so that the latter subsequently (in cosmic time) travelled freely through space, cooling as the universe continued to expand (in cosmic time). The observation of the CMBR therefore confirmed not only the prediction that the Universe was in this hot, dense state early on, but it confirmed the more basic assumption that the universe is three-dimensional and expands through the course of absolute cosmic time.

And if the existence of the CMBR wasn't evidence enough, the best confirmation of the hypothesis that there actually is an absolute cosmic time is really found in its details. Specifically, the otherwise perfect uniformity of the CMBR is expected to have been affected by microscopic vacuum fluctuations prior to its formation, which served as the seeds of cosmic structure. These primary effects should have led to tiny anisotropies within the plasma, and therefore in the radiation that was released when matter became transparent,---the signature of which is supposed to have blown up along with the expansion of space, leading to the primary CMBR anisotropy signature we now observe (\citep{Peebles1970,Sunyaev1970,Zeldovich1972,Weinberg2008}. The consistency of the CMBR anisotropy measurements---described as they are within the framework of the standard cosmological model---with the redshift-distance relation inferred from type Ia supernovae observations, provides the best verification we know of, that the big bang scenario, together with its basic assumption that there is an absolute cosmic time \textit{in Reality}, is correct. 

Through the scientific practice of hypothesis testing, we've found great empirical confirmation that there \textit{is} an absolute Cosmic Time. In fact, if we take the theory that stands behind the CMBR's formation at face value, there is a notable secondary effect in the measured anisotropy, that more directly refutes the commonly-stated belief that Newton's absolute structure can't be observed because we only observe relative motion.\footnote{Cf. \citet[p.~181]{Dainton2001}: ``To account for inertial effects Newton recognized absolute acceleration; having recognized absolute acceleration he also had to recognize absolute velocity\ldots despite the fact that absolute velocity has no physical consequences. In positing absolute velocities, Newtonians expose themselves to the claim that their theory gives rise to empirically indistinguishable states of affairs\ldots. If Newton's theory is correct, there is no empirical test that could be performed that would reveal the answer.'' Dainton recognises neither that there is strong empirical evidence from cosmology to support the assumption of Newtonian absolute structure in cosmology, nor even the fact that the assumption is commonly made in the first place. The latter point is evident in his assertion \citep[p.~295]{Dainton2001}, that ``The most significant assumption made in relativistic cosmology is that the universe is homogeneous and isotropic''; for, relativistic cosmology's assumption that space is always homogeneous and isotropic can't even be made without the \textit{prior assumption} of an absolute cosmic time---which Dainton indeed considers unverifiable structure, and therefore an assumption that's at least as significant as the subsequent one about space. \citet{Maudlin2012}, as well, spends a great deal of effort in removing this ``unobservable structure''.} For indeed, our common inference is that the CMBR's dipole anisotropy is caused by \textit{our own} absolute motion through the Universe, at 370~km/s.

The observation basically undermines the following assertion by \citet{Bell1976}:
\begin{quotation}
The difference of philosophy [between the approaches of Einstein and Lorentz] is this. Since it is experimentally impossible to say which of two uniformly moving systems is \textit{really} at rest, Einstein declares the notions ``really resting'' and ``really moving'' as meaningless.\footnote{``\ldots the phenomena of electrodynamics as well as of mechanics possess no properties corresponding to the idea of absolute rest'' \citep{Einstein1905}.}  For him only the \textit{relative} motion of two or more uniformly moving objects is real. Lorentz, on the other hand, preferred the view that there is indeed a state of \textit{real} rest, defined by the ``aether'', even though the laws of physics conspire to prevent us identifying it experimentally. The facts of physics do not oblige us to accept one philosophy rather than the other.
\end{quotation}
It's simply wrong to assert that we can't determine experimentally whether something is ``really resting'' or ``really moving''---although that's just what \citet{Einstein1905} initially declared. To deny that uniform absolute motion can be experimentally measured, is to deny that the CMBR rest-frame is the absolute rest-frame of our universe,---which is to deny \textit{the} first principle of relativistic cosmology, and therefore everything that comes as a result. In the case of Einstein vs. Einstein---of relative vs. absolute motion---all the great success that's come out of analysing the \textit{one} observable of FLRW big bang cosmology that was actually found \textit{just as expected}---i.e. the CMBR---stands with the assumption of \citet{Einstein1917,Einstein1945}, that there really is an absolute state of rest; and all of cosmology must fall if its most basic assumption is instead supposed to be denied, as \citet{Einstein1905} declared.  Borrowing the words of \citet[p.~104]{Eddington1933}, then, we might state: 
\begin{quote}
The position of absolute time seems impregnable; and if ever relativistic cosmology falls into disrepute cosmic time will be the last stronghold to collapse. \textit{To drop the absolute cosmic time would knock the bottom out of space.}
\end{quote}

There is a truly remarkable point---indeed, more so even than the fact that we can accurately and definitively measure the rate of our own absolute motion, or even the fact that we can detect the residual effects of 13.8 billion-year-old vacuum fluctuations and use them to constrain the parameters of the cosmic expansion that's since taken place---in that, despite living in a pseudo-relativistic, empirically Lorentzian universe that endures absolutely, and despite the fact that we, on Earth, are actually clipping through it at a pretty good pace, we can still make good use, experimentally, of the proper line-element \textit{of privileged observers} (i.e. which is something we \textit{aren't}), simply because the universe is expanding:---

The line-element we use for cosmology doesn't describe our own proper time, the Solar System average, or even that of the Local Group,---but the \textit{absolute cosmic time}. We fully acknowledge that everything is moving through the universe, but note that the Doppler shifts due to this peculiar motion are so small compared to the cosmological redshifts resulting from the propagation of light through expanding space, that both \textit{our own} motion and the motion of \textit{every galaxy we observe}, can realistically be taken as negligible in comparison. Simply because we live in an expanding universe and are capable of observing galaxies at great distances whose light has been highly redshifted as it travelled through expanding space, it turns out that local relativistic effects on the perception of time, simultaneity, etc., due to our own absolute motion through space---that should otherwise be expected to amount to a difference in our perception of ``space'' and ``time''---just don't matter; they're totally insignificant, and the most natural coordinate system for us to use is the ``privileged'' frame instead!

Thus, the objective difference between space and time first suggested within the context of relativistic cosmology by \citet{Einstein1917}, has, after nearly a century, been confirmed at a level of detail that could not have been anticipated, through all of our cosmological redshift measurements and the observed properties of the CMBR. It seems that any hope of returning to a more properly ``relativistic'' description of our Universe, should prove unlikely in the course of progress. Instead, we might aim to learn something about physical reality, and our description of its large scale evolution, from this important result. And this leads us to the last of the four basic assumptions of FLRW cosmology---that the everywhere radially symmetric hypersurfaces of constant cosmic time, given by assumptions i., iii., and iv., should be \textit{orthogonal} to the fundamental world lines; i.e.\ it leads us to assumption ii.\ of FLRW cosmology.

Again (as with just about every aspect of relativistic cosmology), this assumption seems to be based most significantly with Einstein. It's well known that the operational definition of simultaneity that \citet{Einstein1905} wrote down in the first section of his original relativity paper, which comes to mean that synchronous events in special relativity should always be said to have occurred ``simultaneously''---i.e. that ``simultaneity'' and ``synchronicity'' are synonymous\footnote{Cf. \citeapos{Mermin2009} discussion of \citeapos{Einstein1905} postulates of special relativity:
\begin{quote}
\begin{itemize}
\item[$3.$] If a clock at \textit{A} runs synchronously with clocks at \textit{B} and \textit{C}, then the clocks at \textit{B} and \textit{C} also run synchronously relative to each other\ldots
\end{itemize}
The third postulate comes in an obviously equivalent version:
\begin{itemize}
\item[$3'.$] If an event \textit{A} is simultaneous with event \textit{B} and \textit{C}, then events \textit{B} and \textit{C} are also simultaneous.
\end{itemize}
\end{quote}}---is incompatible with the absolute structure of Newtonian mechanics. That's what \citeapos{Einstein1905} original relativity paper comes to mean. Yet it seems that a similar desire, to assume that the absolutely simultaneous events of an FLRW universe should be synchronous in the fundamental cosmic rest-frame, must have contributed to the assumption that the fundamental world lines should be orthogonal to the hypersurfaces of constant cosmic time.

But since the assumption of absolute simultaneity must be made anyway---a fact of nature that all of modern cosmology confirms---it seems both redundant and misguided to add this extra assumption. For one thing, events that are truly simultaneous in these universes won't be synchronous in any frame but the ``privileged'' one anyway, so it's difficult to see what the assumption really buys us. If it's simply supposed to alleviate the difficulty in understanding simultaneity and temporal passage in empirically Lorentzian space-time, when truly simultaneous events aren't synchronous, it should be noted that there is a perfectly natural way to understand this even when making the assumption of absolute time in special relativity \citep{Janzen2012b,Janzen2013}. And finally, it's noteworthy that there really isn't any problem in constructing a homogeneous and isotropic universe with a well defined cosmic time, in which observations appear isotropic to fundamental observers whose world lines \textit{aren't} orthogonal to those slices: the fact was noted already by \citet{deSitter1917b} in the above quotation, where he described ``three-dimensional space ($\mathrm{x}_1, \mathrm{x}_2, \mathrm{x}_3$) with its world-matter as movable in an absolute space''. The world lines of the world-matter in this example aren't orthogonal to the constant cosmic time hypersurfaces, and yet the uniform matter distribution remains uniform, and must appear as such from the perspective of any body that's at rest with the (actually moving) world matter.

In \citeapos{deSitter1917b} example, it could certainly be argued that the ``absolute space'' through which the world-matter is described as ``moving'' is really unnecessary and inconsequential structure, and should therefore be avoided from the point of view of parsimony. It could be argued, e.g., that the world-matter, through its uniform ``motion'', should somehow be the \textit{cause} of the absolute cosmic time. This point, which has been adopted most recently by \citet{Ellis2013}---who claims that the curves ``representing the average motion of matter at each event determined on a suitable averaging scale\ldots will be uniquely determined in any realistic spacetime'', and argues that it is therefore the dynamical matter which should uniquely define ``the present''---would suggest that the fundamental world lines indeed \textit{should be} orthogonal to the cosmic hypersurfaces.

But there are numerous concerns with this point of view. For one thing, although the Universe approximates well as a perfect fluid on the large scale, the world-matter certainly isn't a perfect fluid. Therefore, due to smaller scale inhomogeneities and the fact that no galaxy is actually perfectly at rest with respect to the fundamental rest frame, it hardly seems justified (or helpful) to assume that a global time variable is \textit{caused} by dynamical matter which \textit{nowhere} measures that time as its own. Indeed, arguing for an average causal effect that acts non-locally to define the unique cosmic foliation, as \citet{Ellis2013} does, seems self-contradictory. Actually, his argument, that it's the large scale average motion of matter in the universe that sets the direction of cosmic time, already begs the very point he wants to justify; for in order to say anything at all about a large scale average motion in the universe, one first has to define which spatial slices to call ``the universe'', so that an average can be determined from the world line tangent vectors at each point. But then---even if we choose to neglect the fact that ``instantaneous'' has no meaning if ``time'' isn't defined \textit{a priori},---if cosmic time were \textit{caused} by the average instantaneous velocities of everything in the universe, that causal effect would be maximally non-local, so we run into problems when attempting to state a relativistic cause of cosmic time in any case. 

Thus, there is a strong sense that the argument simply amounts to a chicken-and-egg-type debate: while one might argue that the particular state of existence of everything has a causal influence on the fundamental nature and passage of time, it can just as well be argued that it's incoherent to speak of ``existence'' if there is no passage of time a priori. \citet{Janzen2013} recently argued the latter point in the context of debating \citeapos{Wheeler1990} ``it from bit'' hypothesis. 

Perhaps the best that can be said for the idea, that the average motion of the world-matter should be the fundamental cause of cosmic time, is that it is an attempt to impose a general relativistic cause (world-structure induced by inertia) into an evident aspect of the world (absolute cosmic time) that stands fundamentally opposed to relativity. From this point of view, and considering that the empirical evidence does indicate that there is an absolute cosmic time, it seems that we might rather take a firmer stance on the absolute time assumption, and reject the further restriction that cosmic hypersurfaces---the surfaces describing events of true simultaneity---must necessarily be synchronous in the fundamental rest frame. For while there are cases, like the example mentioned by \citet{deSitter1917b}, in which the assumption amounts to an unobservable difference with FLRW universes in which the fundamental world lines \textit{are} orthogonal to the cosmic hypersurfaces, the same is not true for general expanding universes that are not synchronous in the fundamental rest frame. 

In this case, a solution does exist \citep{Janzen2012a,Janzen2012b} in which the cosmic expansion is \textit{always} due to the presence of a positive cosmological constant, and is therefore physically well defined for all time,---and in which the scale-factor's evolution as a function of the proper time of fundamental observers has, without the need for any exotic matter or an inflationary epoch, \textit{precisely} the same form as it does in a flat $\lambda$CDM universe. This solution challenges the foundational basis (though not the mathematical formalism) of general relativistic dynamics, describing locally observable space-time curvature as a kinematical effect. However, for this very reason, the theory reconciles more naturally with the definition of cosmic time that evidently has to be made, than any attempt to force the concept of a matter-induced absolute time that pervades the whole universe but isn't actually the local time of any of that matter. Indeed, even the horizon problem vanishes when the world-structure is not supposed to be the product of local matter densities. Therefore, rather than forcing unnatural modifications to a theoretical framework that truly seems at-odds with the evidence, which has had to bend to nearly every new empirical finding this past century, this possible solution to the problems of modern cosmology---which provides \textit{exactly} the description that's indicated by all the evidence---challenges us to re-examine the foundations of relativistic cosmology and consider replacing the theory with one that naturally reconciles with all the data.\vspace{-24pt}

\section*{}
{\small Thanks to Leandra Swanner for encouraging this paper to be written, and for useful feedback which helped the argument take its shape. I am indebted to Paul Steinhardt for helping to locate Einstein's relevant correspondences as well as reviewing a draft of this paper, and also to Barbara Wolff at the Einstein Archives for her kind assistance in providing copies of those documents and ensuring that my translations of the German letters were correct. Finally, I thank Guido Bacciagaluppi for feedback on an earlier draft that led to significant improvements.}


\begin{thebibliography}{}

\bibitem[Ade et al.(2013)]{Ade2013} Ade, A.R., et al. 2013. \textit{Planck} 2013 results. XVI. Cosmological parameters. \textit{ArXiv}:1303.5076 [astro-ph.CO]

\bibitem[Alpher et al.(1948)]{Alpher1948a} Alpher, R.A., H.\ Bethe and G.\ Gamow. 1948. The origin of chemical elements. \textit{Phys.\ Rev.}\ \textbf{73}: 803-804

\bibitem[Alpher and Herman(1948)]{Alpher1948b} Alpher, R.A.\ and R.\ Herman. 1948. Evolution of the universe. \textit{Nature} \textbf{162}: 774-775

\bibitem[Alpher and Herman(1950)]{Alpher1950} Alpher, R.A.\ and R.C.\ Herman. 1950. Theory of the origin and relative abundance distribution of the elements. \textit{Rev.\ Mod.\ Phys.}\ \textbf{22}: 153-212

\bibitem[Bacry and L\'{e}vy-Leblond(1968)]{Bacry1968} Bacry, H.\ and J.-M.\ L\'{e}vy-Leblond. 1968. Possible kinematics. \textit{J.\ Math.\ Phys.}\ \textbf{9}: 1605-1614

\bibitem[Bartusiak(2009)]{Bartusiak2009} Bartusiak, M. 2009. \textit{The day we found the universe}. Pantheon, New York

\bibitem[Ben\'{e}treau-Dupin and Smeenk(2013)]{Benetreau2013} Ben\'{e}treau-Dupin, Y.\ and C.\ Smeenk. 2013. What are the foundational issues in cosmology? A report from the Rutgers-UCSC Summer Institute for the Philosophy of Cosmology. Retrieved September 25, 2013 from \url{http://www.rotman.uwo.ca/2013/what-are-the-foundational-issues-in-cosmology-a-report-from-the-rutgers-ucsc-summer-institute-for-the-philosophy-of-cosmology/}

\bibitem[Bell(1976)]{Bell1976} Bell, J.S. 1976. How to teach special relativity. \textit{Progress in Scientific Culture} \textbf{1} (2). Reprinted in Bell, J.S. 1987. \textit{Speakable and Unspeakable in Quantum Mechanics}. Cambridge University Press, Cambridge, 67-80

\bibitem[Bondi(1960)]{Bondi1960} Bondi, H. 1960. \textit{Cosmology}. Cambridge University Press, Cambridge

\bibitem[Bondi and Gold(1948)]{Bondi1948} Bondi, H.\ and T.\ Gold, T. 1948. The steady-state theory of the expanding universe. \textit{Mon.\ Not.\ R.\ Astron.\ Soc.}\ \textbf{108}: 252-270

\bibitem[Callender(2000)]{Callender2000} Callender, C. 2000. Shedding light on time. \textit{Phil.\ Sci.}\ \textbf{67} (3): S587-S599

\bibitem[Carroll et al.(1992)]{Carroll1992} Carroll, S.M., W.H.\ Press and E.L.\ Turner. 1992. The cosmological constant. \textit{Ann.\ Rev.\ Astron.\ Astrophys.}\ \textbf{30}: 499-542

\bibitem[Chandrasekhar(1983)]{Chandrasekhar1983} Chandrasekhar, S. 1983. \textit{Eddington, the most distinguished astrophysicist of his time}. Cambridge University Press, Cambridge

\bibitem[Dainton(2001)]{Dainton2001} Dainton, B. 2001. Time and Space. McGill-Queen’s University Press, Montreal and Kingston

\bibitem[Davis et al.(2007)]{Davis2007} Davis, T.M.\ et al. 2007. Scrutinizing exotic cosmological models using ESSENCE supernova data combined with other cosmological probes. \textit{Astrophys. J.} \textbf{666}: 716-725

\bibitem[de~Sitter(1917a)]{deSitter1917a} de~Sitter, W. 1917a. On the relativity of inertia. Remarks concerning Einstein’s latest hypothesis. \textit{Proc. R. Acad. Amsterdam} 1217-1225

\bibitem[de~Sitter(1917b)]{deSitter1917b} de~Sitter, W. 1917b. Einstein’s theory of gravitation and its astronomical consequences. Third paper. \textit{Mon.\ Not.\ R.\ Astron.\ Soc.}\ \textbf{78}: 3-28

\bibitem[de~Sitter(1930)]{deSitter1930} de~Sitter, W. 1930. On the distances and radial velocities of extra-galactic nebulae, and the explanation of the latter by the relativity theory of inertia. \textit{Proc.\ Nat.\ Acad.\ Sci.}\ \textbf{16}: 474-488

\bibitem[de~Sitter(1931)]{deSitter1931} de~Sitter, W. 1931. Some further computations regarding non-static universes. \textit{Bull. Astronom. Inst. Netherlands} \textbf{223} (6): 141-145

\bibitem[de~Sitter(1933)]{deSitter1933} de~Sitter, W. 1933. The astronomical aspect of the theory of relativity. In \textit{University of California Publications in Mathematics} \textbf{2} (8): 143-196

\bibitem[Dicke(1965)]{Dicke1965} Dicke, R.H.\ et al. 1965. Cosmic black-body radiation.\textit{ Astrophys.\ J.}\ \textbf{142}: 414-419

\bibitem[Dyson(1972)]{Dyson1972} Dyson, F.J. 1972. Missed Opportunities. \textit{Bull.\ Am.\ Math.\ Soc.}\ \textbf{78} (5): 635-652

\bibitem[Eddington(1920)]{Eddington1920} Eddington, A.S. 1920. \textit{Space, Time, and Gravitation}. Cambridge University Press, Cambridge

\bibitem[Eddington(1923)]{Eddington1923} Eddington, A.S. 1923. \textit{The mathematical theory of relativity}, 2nd ed. Cambridge University Press, Cambridge

\bibitem[Eddington(1930)]{Eddington1930} Eddington, A.S. 1930. On the instability of Einstein’s spherical world. \textit{Mon.\ Not.\ R.\ Astron.\ Soc}. \textbf{90}: 668-678

\bibitem[Eddington(1933)]{Eddington1933} Eddington, A.. 1933. \textit{The expanding universe}. Cambridge University Press, Cambridge

\bibitem[Einstein(1905)]{Einstein1905} Einstein, A. 1905. Zur Elektrodynamik bewegter K\"{o}rper. \textit{Ann.\ der Phys.} \textbf{17} (1905). English translation in H.A.\ Lorentz et al., eds. 1952. \textit{The principle of relativity}. Dover Publications, Mineola, New York, 35-65

\bibitem[Einstein(1917)]{Einstein1917} Einstein, A. 1917. Kosmologische betrachtungen zur allgemeinen relativit\"{a}tstheorie. \textit{Sitzungsber.\ K.\ Preuss.\ Akad.\ Wiss}. 142-152. English translation in H.A.\ Lorentz, et al., eds. 1952. \textit{The principle of relativity}. Dover Publications, Mineola, New York, 175-188

\bibitem[Einstein(1923a)]{Einstein1923a} Einstein, A. 1923a. Postcard to H. Weyl. \textit{Einstein Archives} call no.\ 24-081. Post-dated May 23, 1923

\bibitem[Einstein(1922)]{Einstein1922} Einstein, A. 1922. Bemerkung zu der Arbeit von A.\ Friedmann {\glqq}{{\"U}ber} die Kr\"{u}mmung des Raumes''. \textit{Z.\ Phys}. \textbf{11}: 326

\bibitem[Einstein(1923b)]{Einstein1923b} Einstein, A. 1923b. Notiz zu der Arbeit von A.\ Friedmann {\glqq}{{\"U}ber} die Kr\"{u}mmung des Raumes''. \textit{Z.\ Phys}. \textbf{16}: 228

\bibitem[Einstein(1931a)]{Einstein1931a} Einstein, A. 1931a. Zum kosmologischen problem der allgemeinen relativit\"{a}tstheorie. \textit{Sitzungsber.\ Preuss.\ Akad.\ Wiss}. 235-237

\bibitem[Einstein(1931b)]{Einstein1931b} Einstein, A. 1931b. Letter to R.\ Tolman. \textit{Einstein Archives} call no. 23-030. Post-dated June 27, 1931

\bibitem[Einstein(1945)]{Einstein1945} Einstein, A. 1945. \textit{The Meaning of Relativity}, 2nd ed. Princeton University Press, Princeton

\bibitem[Einstein(1949)]{Einstein1949} Einstein, A. 1949. Remarks to the essays appearing in this volume. In P.A.\ Schilpp, ed. \textit{Albert Einstein: Philosopher-Scientist}. Open Court Publishing Company, Peru, Illinios, 663-688

\bibitem[Einstein and de~Sitter(1932)]{Einstein1932} Einstein, A.\ and W.\ de~Sitter. 1932. On the relation between the expansion and the mean density of the universe. \textit{Proc.\ Nat.\ Acad.\ Sci.} \textbf{18} (3): 213-214

\bibitem[Ellis(2003)]{Ellis2003} Ellis, G.F.R. 2003. A historical review of how the cosmological constant has fared in general relativity and cosmology. \textit{Chaos, Solitons and Fractals} \textbf{16}: 505-512

\bibitem[Eillis(2013)]{Ellis2013} Ellis, G.F.R. 2013. The arrow of time and the nature of spacetime. Studies in the History and Philosophy of Modern Physics \textbf{44}: 242-262

\bibitem[Ellis and Krasi\'{n}ski(1999)]{Ellis1999} Ellis, G.F.R.\ and A. Krasi\'{n}ski. 1999. Editors note: On the curvature of space and on the possibility of a world with constant negative curvature of space. \textit{Gen.\ Rel.\ Gravitat.} \textbf{31} 1985-1990

\bibitem[Farrell(2005)]{Farrell2005} Farrell, J. 2005. \textit{The day without yesterday: Lema\^{i}tre, Einstein, and the birth of modern cosmology}. Thunder’s Mouth Press, New York

\bibitem[Forscher(1963)]{Forscher1963} Forscher, B.K. 1963. Chaos in the brickyard. Science \textbf{142} (3590): 339

\bibitem[Friedmann(1922)]{Friedmann1922} Friedman, A.: \"{U}ber die Kr\"{u}mmung des Raumes. \textit{Z.\ Phys}. \textbf{10}: 377-386. English translation: Friedman, A. On the curvature of space. \textit{Gen.\ Rel.\ Gravitat}. \textbf{31}: 1991-2000
 
\bibitem[Friedmann(1924)]{Friedmann1924} Friedmann, A. 1924. \"{U}ber die M\"{o}glichkeit einer Welt mit konstanter negativer Kr\"{u}mmung des Raumes. \textit{Z.\ Phys}. \textbf{21}: 326-332. English translation: Friedmann, A. On the possibility of a world with constant negative curvature of space. \textit{Gen.\ Rel.\ Gravitat.} \textbf{31} 2001-2008

\bibitem[Gamow(1946)]{Gamow1946} Gamow, G. 1946. Expanding universe and the origin of elements. \textit{Phys.\ Rev.} \textbf{70}: 572-573

\bibitem[Gamow(1948)]{Gamow1948} Gamow, G. 1948. The evolution of the universe. \textit{Nature} \textbf{162}: 680-682

\bibitem[Gamow(1970)]{Gamow1970} Gamow, G. 1970. My world line: an informal autobiography. Viking Press, New York

\bibitem[Goobar and Perlmutter(1995)]{Goobar1995} Goobar, A. and S.\ Perlmutter. 1995. Feasibility of measuring the cosmological constant $\Lambda$ and mass density $\Omega$ using type Ia supernovae. \textit{Astrophys.\ J.} \textbf{450}: 14-18

\bibitem[Gribbin(1986)]{Gribbin1986} Gribbin, J. 1986. \textit{In search of the big bang}. Bantam Books, New York

\bibitem[Guth(1981)]{Guth1981} Guth, A.H. 1981. Inflationary universe: A possible solution to the horizon and flatness problems. \textit{Phys. Rev. D} \textbf{23} (2) 347-356

\bibitem[Hawking(1967)]{Hawking1967} Hawking, S.W. 1967. The occurrence of singularities in cosmology. III. Causality and singularities. \textit{Proc.\ R.\ Soc.\ London A, Math.\ Phys.\ Sci.} \textbf{300} (1461): 187-201

\bibitem[Heckmann(1931)]{Heckmann1931} Heckmann, O. 1931. \"{U}ber die Metrik des sich ausehenden Universums. \textit{Ver\"{o}ffentlichungen der Universit\"{a}ts-sternwarte zu G\"{o}ttingen} \textbf{2}: 127-131

\bibitem[Hicken et al.(2009)]{Hicken2009} Hicken, M.\ et al. 2009. Improved dark energy constraints from ~100 new CfA supernova type Ia light curves. \textit{Astrophys.\ J}. \textbf{700}: 1097-1140 

\bibitem[Hinshaw(2013)]{Hinshaw2013} Hinshaw, D.\ et al. 2013. Nine-year \textit{Wilkinson Microwave Anisotropy Probe} (\textit{WMAP}) observations: cosmological parameter results. \textit{Astrophys. J. Supplement Series} \textbf{208} (19): 1-25

\bibitem[Hoyle(1948)]{Hoyle1948} Hoyle, F. 1948. A new model for the expanding universe. \textit{Mon.\ Not.\ R.\ Astron.\ Soc}. \textbf{108}: 372-382

\bibitem[Hubble(1929)]{Hubble1929} Hubble, E. 1929. A relation between distance and radial velocity among extra-galactic nebulae. \textit{Proc.\ Nat.\ Acad.\ Sci}. \textbf{15}: 168-173

\bibitem[Ijjas(2013)]{Ijjas2013} Ijjas, A., P.J.\ Steinhardt and A.\ Loeb. 2013. Inflationary paradigm in trouble after Planck2013. \textit{Phys.\ Lett.\ B} \textbf{723}: 261-266

\bibitem[Janzen(2012a)]{Janzen2012a} Janzen, D. 2012a. A solution to the cosmological problem of relativity theory. PhD thesis. University of Saskatchewan. \url{http://hdl.handle.net/10388/ETD-2012-03-384}

\bibitem[Janzen(2012b)]{Janzen2012b} Janzen, D. 2012b. A critical look at the standard cosmological picture. Fourth Prize entry in FQXi’s Essay Contest, \textit{Questioning the Foundations}. arXiv:1303.2549

\bibitem[Janzen(2013)]{Janzen2013} Janzen, D. 2013. Time is the denominator of existence, and bits come to be in it. Submitted to FQXi’s 2013 Essay Contest, \textit{It From Bit or Bit From It?}

\bibitem[Kardashev(1967)]{Kardashev1967} Kardashev, N. 1967. Lema\^{i}tre's universe and observations. \textit{Astrophys.\ J.\ Lett}. \textbf{150}: L135-L139

\bibitem[Kolb(2007)]{Kolb2007} Kolb, R. 2007. The greatest discovery Einstein didn’t make. In Brockman, J.\ ed.\ \textit{My Einstein: essays by twenty-four of the world's leading thinkers on the man, his work, and his legacy}. Pantheon, New York 

\bibitem[Kragh(2003)]{Kragh2003} Kragh, H. 2003. \textit{Cosmology and Controversy: the historical development of two theories of the universe}. Princeton University Press, Princeton

\bibitem[Lema\^{i}tre(1925)]{Lemaitre1925} Lema\^{i}tre, G. 1925. Note on de~Sitter’s universe.\textit{ J.\ Math.\ Phys}. \textbf{4}: 188-192

\bibitem[Lema\^{i}tre(1927)]{Lemaitre1927} Lema\^{i}tre, G. 1927. Un Univers homog\`{e}ne de masse constante et de rayon croissant rendant compte de la vitesse radiale des nébuleuses extra-galactiques. \textit{Annales de la Soci\'{e}t\'{e} Scientifique de Bruxelles A}, \textbf{47}: 49-59. English translation: Lema\^{i}tre, G. 1931. A Homogeneous Universe of Constant Mass and Increasing Radius accounting for the Radial Velocity of Extra-Galactic Nebulae. \textit{Mon.\ Not.\ R.\ Astron.\ Soc}. \textbf{91}: 483-490

\bibitem[Lema\^{i}tre(1933)]{Lemaitre1933} Lema\^{i}tre, G. 1933. L'Univers en expansion. \textit{Annales de la Soci\'{e}t\'{e} Scietifique de Bruxelles} \textbf{53}: 51-85 (1933). English translation: Lema\^{i}tre, G. 1997. The expanding universe. \textit{Gen.\ Rel.\ Gravitat}. \textbf{29}: 641-680

\bibitem[Lema\^{i}tre(1949a)]{Lemaitre1949a} Lema\^{i}tre, G. 1949a. The cosmological constant. In P.\ A.\ Schilpp, ed. \textit{Albert Einstein: Philosopher-Scientist}. Open Court Publishing Company, Peru, Illinios, 437-456

\bibitem[Lema\^{i}tre(1949b)]{Lemaitre1949b} Lema\^{i}tre, G. 1949b. Cosmological Application of Relativity. \textit{Rev.\ Mod.\ Phys}. \textbf{21}: 357-366

\bibitem[Maudlin(2012)]{Maudlin2012} Maudlin, T. 2012. \textit{Philosophy of Physics: Space and Time.} Princeton University Press, Princeton

\bibitem[McVittie (1967)]{McVittie1967} McVittie, G.C. 1967. Georges Lemaitre (obituary notice). \textit{Quarterly J. R. Astron. Soc.} \textbf{8}: 294-297

\bibitem[Mermin(2009)]{Mermin2009} Mermin, N.D. 2009. Plane geometry in spacetime. In W.C.\ Myrvold and J.\ Christian, eds. \textit{Quantum reality, relativistic causality, and closing the epistemic circle}. Springer, 327-347

\bibitem[Pauli(1958)]{Pauli1958} Pauli, W. 1958. Theory of relativity. Pergamon Press, New York
 
\bibitem[Peebles and Ratra(2003)]{Peebles2003} Peebles, P.J.E.\ and B.\ Ratra. 2003. The cosmological constant and dark energy. \textit{Rev. Mod. Phys}. \textbf{75} (2): 559-606

\bibitem[Peebles(1970)]{Peebles1970} Peebles, P.J.E.\ and J.T.\ Yu. 1970. Primeval adiabatic perturbation in an expanding universe. \textit{Astrophys.\ J.} \textbf{162}: 815-836

\bibitem[Penrose(2005)]{Penrose2005} Penrose, R. 2005. \textit{The Road to Reality}. Vintage, London

\bibitem[Penzias and Wilson(1965)]{Penzias1965} Penzias, A.A.\ and R.W.\ Wilson. 1965. A measurement of excess antenna temperature at 4080 MC/s. \textit{Astrophys. J}. \textbf{142} 419-421

\bibitem[Perlmutter et al.(1999)]{Perlmutter1999} Perlmutter, S.\ et al. 1999. Measurements of $\Omega$ and $\Lambda$ from 42 high-redshift supernovae. \textit{Astrophys.\ J}. \textbf{517}: 565-586

\bibitem[Petrosian et al.(1967)]{Petrosian1967} Petrosian, V., E.\ Salpeter and P.\ Szekeres. 1967. Quasi-stellar objects in universes with non-zero cosmological constant. \textit{Astrophys. J}. \textbf{147} 1222-1226

\bibitem[Putnam(1967)]{Putnam1967} Putnam, H. 1967. Time and physical geometry. \textit{J.\ Phil.} \textbf{64} (8): 240-247

\bibitem[RAS(1930)]{RAS1930} RAS. 1930. Meeting of the Royal Astronomical Society. Friday, 1930 May 9. \textit{The Observatory} \textbf{53}: 161-164

\bibitem[Riess et al.(1998)]{Riess1998} Riess, A.G.\ et al. 1998. Observational evidence from supernovae for an accelerating universe and a cosmological constant. \textit{Astronom.\ J}. \textbf{116}: 1009-1038

\bibitem[Riess et al.(2004)]{Riess2004} Riess, A.G.\ et al. 2004. Type Ia supernovae discoveries at $z > 1$ from the \textit{Hubble Space Telescope}: evidence for past deceleration and constraints on dark energy evolution. \textit{Astrophys.\ J}. \textbf{607}: 665-687

\bibitem[Riess et al.(2007)]{Riess2007} Riess, A.G.\ et al. 2007. New \textit{Hubble Space Telescope} discoveries of type Ia supernovae at $z \geq 1$: narrowing constraints on the early behavior of dark energy. \textit{Astrophys.\ J}. \textbf{659}: 98-121

\bibitem[Riess et al.(2011)]{Riess2011} Riess, A.G.\ et al. 2011. A 3\% solution: determination of the Hubble constant with the \textit{Hubble Space Telescope} and Wide Field Camera 3. \textit{Astroph. J}. \textbf{730} (119): 1-18

\bibitem[Robertson(1928)]{Robertson1928} Robertson, H.P.: On relativistic cosmology. \textit{Phil. Mag}. \textbf{5}: 835-848

\bibitem[Robertson(1929)]{Robertson1929} Robertson, H.P. 1929. On the foundations of relativistic cosmology. \textit{Proc. Nat. Acad. Sci}. \textbf{15}: 822-829

\bibitem[Robertson(1933)]{Robertson1933} Robertson, H.P. 1933. Relativistic cosmology. \textit{Rev.\ Mod.\ Phys}. \textbf{5} (1): 62-90

\bibitem[Sandage(1958)]{Sandage1958} Sandage, A. 1958. Current problems in the extragalactic distance scale. \textit{Astrophys. J}. \textbf{127} (3): 513-526

\bibitem[Shklovsky(1967)]{Shklovsky1967} Shklovsky, J. 1967 On the nature of ``standard'' absorption spectrum of the quasi-stellar objects. \textit{Astrophys. J. Lett}. \textbf{150}: L1-L3

\bibitem[Slipher(1913)]{Slipher1913} Slipher, V.M.: The radial velocity of the Andromeda Nebula. Lowell Observatory Bulletin \textbf{2}: 56-57

\bibitem[Smith(1990)]{Smith1990} Smith, R.W. 1990. Edwin P.\ Hubble and the transformation of cosmology. \textit{Physics Today} \textbf{43}: 52-58

\bibitem[Steinhardt(2011)]{Steinhardt2011} Steinhardt, P.J. 2011. The inflation debate. \textit{Sci. Am}. \textbf{304} (4): 36-43

\bibitem[Steinhardt and Turok(2007)]{Steinhardt2007} Steinhardt, P.J.\ and N.\ Turok. 2007. Endless universe: beyond the big bang. Doubleday, New York

\bibitem[Sunyaev and Zel'dovich(1970)]{Sunyaev1970} Sunyaev, R.A.\ and Y.B.\ Zel'dovich. 1970. The interaction of matter and radiation in the hot model of the universe, II. \textit{Astrophysics and Space Science} \textbf{7} 20-30

\bibitem[Suzuki et al.(2012)]{Suzuki2012} Suzuki, N.\ et al. 2012. The \textit{Hubble Space Telescope} cluster supernova survey. V. Improving the dark-energy constraints above $z > 1$ and building an early-type-hosted supernova sample. \textit{Astrophys. J}. \textbf{746} (85): 1-24

\bibitem[Tolman(1931a)]{Tolman1931a} Tolman, R.C. 1931a. Letter to A.\ Einstein. \textit{Einstein Archives} call no. 23-031. Post-dated September 14, 1931

\bibitem[Tolman(1931b)]{Tolman1931b} Tolman, R.C. 1931b. On the theoretical requirements for a periodic behaviour of the universe. \textit{Phys. Rev}. \textbf{38}: 1758-1771

\bibitem[Weinberg(1989)]{Weinberg1989} Weinberg, S. 1989. The cosmological constant problem. \textit{Rev. Mod. Phys}. \textbf{61} (1): 1-23

\bibitem[Weinberg(2008)]{Weinberg2008} Weinberg, S. 2008. \textit{Cosmology}. Oxford University Press, Oxford

\bibitem[Weyl(1923a)]{Weyl1923a} Weyl, H. 1923a. Raum, zeit, materie: vorlesungen \"{u}ber allgemeine relativit\"{a}tstheorie, f\"{u}nfte umgearbeitete auflage. Julius Springer, Berlin

\bibitem[Weyl(1923b)]{Weyl1923b} Weyl, H. 1923b. Zur allgemeinen relativit\"{a}tstheorie. \textit{Phys. Z.} \textbf{24}: 230-232 (1923b). English translation: Weyl, H. 2009. Republication of: On the general relativity theory. \textit{Gen.\ Relativ.\ Gravitat}. \textbf{35}: 1661-1666

\bibitem[Weyl(1924)]{Weyl1924} Weyl, H. 1924. Massentr\"{a}gheit und kosmos. \textit{Naturwissenschaften} \textbf{12}: 197-204

\bibitem[Weyl(1930)]{Weyl1930} Weyl, H. 1930. Redshift and relativistic cosmology. \textit{Phil. Mag}. \textbf{9}: 936-943

\bibitem[Wheeler(1990)]{Wheeler1990} Wheeler, J.A. 1990. Information, physics, quantum: the search for links. In W.\ Zurek, ed.: \textit{Complexity, entropy, and the physics of information}. Addison-Wesley, Boston

\bibitem[Zel'dovich(1972)]{Zeldovich1972} Zel'dovich, Y.B. 1972. A hypothesis, unifying the structure and the entropy of the Universe. \textit{Mon.\ Not.\ R.\ Astron.\ Soc}. \textbf{160}: 1P-3P

\end{thebibliography}
\end{document}